\documentclass[preprint,5p,sort&compress,times,twocolumn]{elsarticle}

\usepackage[utf8]{inputenc}
\usepackage{physics} 
\usepackage{mhchem} 
\usepackage{nicefrac} 
\usepackage{appendix} 
\usepackage{amsmath,mathtools}
\usepackage{graphicx,tikz, pgfplots, subfigure}
\pgfplotsset{compat=1.16}
\pgfplotsset{scaled y ticks=false}
\usepackage{xcolor}
\usepackage{hyperref} 
\usepackage{multirow} 
\usepackage[font=small,labelfont=bf,
   justification=justified,
   format=plain]{caption} 

\journal{Chinese Journal of Physics}

\begin{document}

\begin{frontmatter}

\title{Spin-Flavor Precession Phase Effects in Supernova}

\author[first]{T.~Bulmus}
\ead{taygun.bulmus@msgsu.edu.tr}
\author[first]{Y.~Pehlivan}
\affiliation[first]{organization={Mimar Sinan Fine Arts University},
            addressline={Sisli}, 
            city={Istanbul},
            postcode={34380}, 
            country={Türkiye}}

\begin{abstract}
We study the phase effects driven by neutrino magnetic moment for Majorana
neutrinos in a core collapse supernova. A neutrino with a large magnetic
moment is emitted in a superposition of energy eigenstates from the
neutrinosphere. These energy eigenstates can interfere to create a phase effect
at a partially adiabatic spin flavor precession (SFP) resonance.  We examine the
dependence of the SFP phase effect on the size of the neutrino magnetic moment
as well as its variation with the post-bounce time. In particular, at late
post-bounce times  the SFP resonance becomes wider and eventually overlaps with
the Mikheev-Smirnov-Wolfenstein (MSW) resonance. At this point Landau-Zener
criteria for adiabaticity can no longer be applied to individual 
resonances, but we show that SFP phase effect is still present after
the overlap. We also discuss the observability of the SFP phase effect 
at Deep Underground Neutrino Experiment (DUNE). Our analysis reveals that at
low energies event rates do not fluctuate despite the presence of a sizable SFP
phase effect. We find larger event rate fluctuations at high energies, but these
fluctuations are also erased in the energy spectra of the observed charged
leptons. A more refined treatment of electron fraction and the inclusion of
neutrino-neutrino interactions may change our conclusions for observability in
future studies. 
\end{abstract}


\end{frontmatter}




\section{Introduction}
\label{sec:INTRODUCTION}

Neutrino's anomalous magnetic moment causes its spin to precess around a
magnetic field \cite{Pauli:1941zz, Lee:1977tib, SHROCK1982359}. This, coupled
with the ordinary flavor evolution in vacuum, gives rise to the SFP phenomenon
\cite{Fujikawa:1980yx}. Ordinarily, the Standard Model predicts the neutrino
magnetic moment to be of the order of $10^{-20} \mu_B$ where $\mu_B$ denotes the
Bohr magneton \cite{Fujikawa:1980yx, Balantekin:2013sda}. This value is too small to
be consequential in most settings, but it is also possible that the neutrino
magnetic moment is larger than the Standard Model prediction \cite{Bell:2005kz,
Babu:2020ivd}. Current experimental upper bound for neutrino magnetic moment is
of the order of $10^{-11} \mu_B$ \cite{Zyla:2020zbs}. However, somewhat stronger
bounds can be provided with astrophysical arguments \cite{Raffelt:1999gv}. For a
recent review of neutrino electromagnetic properties, see Ref.
\cite{Giunti:2014ixa}. Here, we assume that the neutrino magnetic moment is of
the order of $10^{-16} \mu_B$ or larger.

If the magnetic field is perpendicular to the velocity of the neutrino, then
spin precession causes neutrino's helicity to oscillate. As a result, SFP is
modified in matter because positive and negative helicity neutrinos interact
differently \cite{Voloshin:1986ty, Okun:1986na, Lim:1987tk, Akhmedov:1988uk}.
In particular, each interaction forces the neutrino back into a flavor state so
that SFP is suppressed in high matter density. But, under the right conditions,
the effects of the matter interactions and the vacuum oscillations can cancel
each other.  Around this cancellation region even a relatively weak magnetic
field can cause significant helicity transformation, which is known as the SFP
resonance \cite{Lim:1987tk, Akhmedov:1988uk}. It is analogous to the MSW
resonance which happens when neutrinos undergo ordinary flavor oscillations in a
medium and a similar cancellation leads to significant flavor transformation
even for a very small mixing angle \cite{Wolfenstein:1977ue, Mikheev:1986wj,
1986PhRvL..57.1275P}. 

In this paper, we consider the phase effects driven by the neutrino magnetic
moment and the associated SFP resonance in a core collapse supernova. We call
this the \emph{SFP phase effect} in order to distinguish it from the
\emph{ordinary phase effect} \cite{Haxton:1986bc, Mikheev:1987wa, Kuo:1989qe}
associated with the standard flavor oscillations and the MSW resonances. The
appearance of ordinary phase effect in a core collapse supernova is pointed out
in many references \cite{Fogli:2003dw, Kneller:2005hf, Kneller:2007kg,
Galais:2009wi, Kneller:2010sc, Horiuchi:2018ofe} and examined in detail in Ref.
\cite{Dasgupta:2005wn}. Both the standart flavor oscillations and SFP can take
plase without the phase effects, whose development require two additional
conditions to be satisfied in the given order: First, neutrinos should be
evolving in superpositions of energy eigenstates, and second, they should pass
through a partially adiabatic resonance. Partial adiabaticity is described by
the Landau-Zener jumping probability \cite{1932PhyZS...2...46L,
1932RSPSA.137..696Z, 1981PhRvA..23.3107R, Kuo:1989qe, Smirnov:2004zv}, and a
Stoke's phase \cite{PhysRevA.50.843, PhysRevA.55.R2495}. It is the sum of the
Stoke's phase and the relative phase acquired by the energy eigenstates during
their evolution that creates the phase effects.  Both the ordinary phase effect
and the SFP phase effect emerge through the same underlying mechanism, but they
differ in at least two important aspects to be discussed in what follows.

The ordinary phase effect appears when neutrinos go through two subsequent
partially adiabatic MSW resonances in the same channel, which is often seen when
the shock wave in a supernova creates a sharp local dip in matter density. In
the usual scenario, the neutrino is produced nearly in an energy eigenstate at
the center of the supernova. As it propagates, the first partially adiabatic MSW
resonance transforms the neutrino into a superposition of two energy eigenstates
and the second one creates the phase effect through an interference between
them. In contrast, the SFP phase effect can appear even if there is only one
partially adiabatic SFP resonance because, assuming it has a large magnetic
moment, the neutrino is already produced in a superposition of energy
eigenstates at the center. This is due to the fact that near the proto-neutron
star, both the matter density and the magnetic field are expected to be high
\cite{2002RvMP...74.1015W, Kotake:2005zn, Janka:2006fh}. The strong magnetic
field pulls the energy eigenstates close to the spin projection eigenstates
along itself while the high matter density pulls them close to flavor
eigenstates. As a result, a neutrino produced with a particular flavor and
helicity is necessarily born into a superposition of energy eigenstates, which
can subsequently interfere at the first partially adiabatic resonance. 

Consequently, the first important difference between the ordinary phase effect
and the SFP phase effect is in their ubiquity: The ordinary phase effect is
limited to the part of the neutrino spectrum which experiences two partially
adiabatic resonances in the same channel.  The particulars of the shock wave
propagation as well as the energy dependence of the adiabaticity parameter for
the MSW resonances determine which part of the spectrum is effected.  This can
be seen, for example, in Fig. 2 of Ref.  \cite{Fogli:2003dw}, Fig. 7 of Ref.
\cite{Kneller:2005hf}, or Figs. 17-21 of Ref. \cite{Kneller:2007kg}. In
contrast, if it is present, the SFP phase effect affects the entire neutrino
energy spectrum because a single SFP resonance is sufficient and its adiabaticity
does not depend on the neutrino energy. 

The second important difference between SFP phase effect and the ordinary phase
effect involves the non-universal nature of the SFP resonance. MSW resonance is
universal in the sense that its width in density scale is fixed by the vacuum
mixing parameters. In contrast, the width of the SFP resonance depends on the
way the magnetic field varies with respect to density. At late post-bounce
times, it is possible that the SFP resonance becomes wide enough to overlap with
the subsequent MSW resonance. Such an overlap breaks down the Landau-Zener
formulation since it is only applicable to isolated resonances. We demonstrate
that, after the overlap, the SFP phase effect continues to be present while both
resonances appear to become adiabatic from a naive application of the
Landau-Zener formulation. 

Phase effects cause the neutrino survival and transition probabilities to
oscillate with energy. These oscillations typically occur on very small scales
so that the results may look randomized if the energy binning can not resolve
them. In most cases, one eliminates the phase effects by averaging them out but
there are also cases where it is important to understand them. For example,
while discussing the impact of turbulence on neutrino oscillations, Ref.
\cite{Kneller:2010sc} points out to the interplay between turbulence   and phase
effects. In particular, it is shown that the impact of the turbulence can be
numerically identified only when it starts to dominate over the phase effects,
i.e., once it starts to create correlations in shorter energy scales than the
phase effects. 

The reason for the fast variation with energy is easy to understand in the case
of non-overlapping resonances: According to the Landau-Zener approach, for each
resonance channel the energy eigenstates can be considered pairwise and hence
there is only one relevant phase difference in each case. Since this phase difference is
typically accumulated over a long distance (from production to the
resonance or between two resonances) it is large and hence vary
fast with energy. However, the picture is less clear when the SFP resonance wide
enough to overlap with the MSW resonance. In this case, at least three energy
eigenstates mix at the same time and hence there are two phase differences
involved in the problem. Depending on the amount of overlap between the
resonances, the accumulation distance can be shorter for at least one of these
relative phases. Since an analytical treatment of the problem is challenging, it
is difficult to make a purely theoretical deliberation at this point. For the
model that we study, we find that the phase effects continue to be effective in
small energy scales in the case of a partial overlap.  However, we believe that
this problem deserves to be investigated more systematically than we attempt in
this paper. 

From an observational point of view, capturing phase effects is challenging due
to the finite detector energy resolutions. Ref. \cite{Dasgupta:2005wn} concluded
that the ordinary phase effect cannot be observed with the current detection
capabilities. While we primarily focus on understanding the SFP phase effect, we
also touch upon the possibility of observing it. We argue that when the
oscillations of neutrino survival and transition probabilities occur on a small
energy scale, it is more appropriate to think of them as random fluctuations
within certain bounds because relative phases are also sensitive to small
variations in external conditions. For this reason we treat the SFP phase effect
as an \emph{uncertainty} within which neutrino survival and transition
probabilities can randomly fluctuate. Looking at the problem in this way leads
to the conclusion that, irrespective of the detector energy resolution, phase
effects cannot be observed if the neutrino flux is very high, e.g., at early
post-bounce times. This is because if the detector captures many neutrinos in
the same energy bin, then the random fluctuations would cancel each other in
each one of them. On the other hand, if only a few neutrinos
are captured in each energy bin as would be the case at late post-bounce times,
then this cancellation may be partial and some residual randomization can
remain. However, the latter can still be wiped out with imperfect detector
energy resolution.  

The technical details of our paper are as follows: We assume that neutrinos are
Majorana particles and derive explicit analytical formulas to calculate the size
of the uncertainties in survival and transition probabilities. This analytical
treatment is valid for any density and magnetic field distribution as long as
SFP and MSW resonances are decoupled in the sense that they do not overlap. In
order to highlight a few basic features of the SFP phase effect, we first run
illustrative simulations using an exponentially decreasing matter density and a
magnetic field which decreases with the square of the distance from the center.
In these illustrative simulations we mimic the effect of a shock wave by
decreasing the central density with the post-bounce time. We show that the
decoupling approximation (and our analytical treatment with it) fails at later
post-bounce times due to the spreading of SFP resonance with dropping central
density. This is especially true if the neutrino magnetic moment is large or if
the magnetic field is strong. We also run simulations with a realistic
density distribution based on a $6M_{\odot}$ helium core presupernova model
\cite{1987ESOC...26..325N} and a parametric shock wave \cite{Fogli:2003dw} to
demonstrate how the SFP phase effect may affect the survival and transition
probabilities of neutrinos arriving Earth. 

We start by reviewing the oscillations, precession and interactions of
Majorana neutrinos inside the supernova in Section II. In Section III, we discuss SFP and MSW
resonances, particularly focusing on the conditions under which they can be
treated as two separate two-level problems. In Section IV, we discuss the
adiabatic evolution of neutrinos inside the supernova, and their subsequent
decoherence on their way to Earth. In Section V, we consider the hypothetical
case of the zero vacuum mixing angle in 
order to remove the MSW resonance from the picture and focus
on the phase effect between the production point and the SFP resonance. In
Section VI, we consider the non-zero mixing angle case and thus include the MSW
resonances. In Section VII, we discuss the observability of the SFP phase effect
at late post-bounce times.  In Section VIII, we present our discussion and
conclusions.

\section{Neutrinos inside the supernova}
\label{sec:systemDynamics}

We work in the effective two flavor mixing scheme with 1-3 mixing parameters
which is the relevant part of the full mixing parameter space for supernova.
See, for example, Ref. \cite{Ahriche:2003wt} for three flavor effects. We
denote the negative helicity flavor degrees of freedom by $\ket{\nu_e}$ and
$\ket{\nu_x}$. They respectively correspond to the electron flavor and an
orthogonal flavor combination. Corresponding positive helicity states are
respectively denoted by $\ket{\bar\nu_e}$ and $\ket{\bar\nu_x}$. Majorana
neutrinos are their own antiparticles but, as far as their production and
detection are concerned, a positive helicity Majorana neutrino behaves very
similar to a positive helicity Dirac antineutrino. For this reason, it is
conventional to refer to Majorana neutrinos with positive helicities as
antineutrinos. We also adopt this convention, but we refer to all degrees of
freedom as neutrinos when no distinction is necessary.

A magnetic field turns a Majorana neutrino into a Majorana antineutrino by
flipping its helicity. The resulting effect is described by the
Hamiltonian\footnote{We use outer products rather than matricies
because writing down $4\times 4$ matrices is
impractical in the two-column format.} 
\begin{equation}
\label{Hmu}
H_{\mu}\mkern-1mu(r)\!=\! \mu B(r)
\qty(\ket{\nu_{e}}\bra{\bar\nu_x}\!+\!\ket{\bar\nu_x}\bra{\nu_e}\!-\!\ket{\nu_x}\bra{\bar\nu_e}
\!-\!\ket{\bar\nu_e}\bra{\nu_x}). 
\end{equation}
This Hamiltonian only includes flavor off-diagonal terms because flavor-diagonal
terms vanish identically due to the reality condition of the Majorana spinors
\cite{Giunti:2014ixa, Pehlivan:2014zua}. $B(r)$ in Eq. (\ref{Hmu}) denotes the component of the
magnetic field perpendicular to the neutrino's direction of motion. For those
neutrinos to be detected in experiments, it depends on the orientation of the
supernova with respect to Earth in addition to the particulars of the supernova
dynamics. For this reason, it is difficult to be specific about $B(r)$.
Moreover, it is always the $\mu B(r)$ combination which appears in equations. We
find that, for the purposes of this paper the important parameters are the
values of $\mu B$ on the surface of the neutrinosphere and around the SFP
resonance region. In particular, how the magnetic field changes in between these
two points is less important. For definiteness, we use a magnetic field which
decreases with the distance $r$ from the center of the supernova as
\cite{deGouvea:2012hg, deGouvea:2013zp, Kharlanov:2020cti, Sasaki:2021bvu}
\begin{equation}
\label{B value}
B(r)=B_0\left(\frac{r_{\mbox{\footnotesize{mag}}}}{r}\right)^2 
\end{equation}
where $r_{\mbox{\footnotesize{mag}}}=50$ km. Neutrino flavor evolution starts
from the surface of the proto-neutron star which we also take to be at $R=50$
km. Therefore, $B_0$ is effectively the magnetic field on the surface of the
neutrinosphere. Its value can range from a conservative $10^{12} \text{G}$ to
extreme values such as $10^{16} \text{G}$ \cite{Kotake:2005zn, Sawai:2005pr}.
For example, taking $B_0=10^{15} \text{G}$ and $\mu=3\times 10^{-16} \mu_B$
leads to an energy separation of
\begin{equation}
\label{muB value}
\mu B_0=1.7\times 10^{-9} \mbox{ eV}
\end{equation}
between the helicity eigenstates on the surface of the neutrinosphere. This is
larger than the typical separation of $10^{-11}$ eV between neutrino mass
eigenstates in vacuum but smaller than the typical separation of $10^{-8}$ eV
between the flavor eigenstates in matter. We discuss these figures below. 

Neutrinos interact with other particles and with each other in the supernova
\cite{Flowers:1976kb, fuller&mayle}. Here, we ignore the neutrino-neutrino
interactions\footnote{Neutrino-neutrino interactions turn the neutrinos
streaming out of a supernova into a self interacting many-body system
\cite{Pantaleone:1992eq, Pantaleone:1992xh} with non-linear behavior. See Ref.
\cite{Duan:2010bg, Chakraborty:2016yeg, Volpe:2023met} for review. Without a
large neutrino magnetic moment, the effects of the neutrino-neutrino
interactions on neutrino flavor evolution is extensively studied.  For normal
mass hierarchy (NH) these interactions generally have minimal effect on the
flavor evolution.  For inverted mass hierarchy (IH) they cause different flavors
to swap parts of their energy spectra with each other in the first few hundred
km. With a large neutrino magnetic moment, the possible effects are more
complicated. See, for example, Refs. \cite{deGouvea:2012hg, deGouvea:2013zp,
Pehlivan:2014zua, Kharlanov:2020cti, Sasaki:2021bvu}. We comment on how they can
affect our results in the Conclusions.}. The interactions with the other
particles can be further limited to the forward scattering alone because these
are the only terms which add up coherently \cite{Wolfenstein:1977ue}. At the MeV
energy scale relevant for the supernova, all weak interactions can be treated
within the Fermi four point model in terms of the Fermi coupling constant $G_F$.
With these considerations, the Hamiltonian describing the vacuum oscillations
and interactions in an unpolarized neutral medium can be written as
\begin{equation}
\label{Hnunu}
\begin{split}
H_{\nu \leftrightarrow \nu}(r)
=&\qty(-\tfrac{\delta m^2}{2E}\cos{2\theta}+\tfrac{\sqrt{2}G_Fn(r)}{m_n}\tfrac{3Y_e-1}{2}) 
\ket{\nu_e}\bra{\nu_e}\\ 
+&\qty(\tfrac{\delta m^2}{2E_\nu} \cos{2\theta}-\tfrac{\sqrt{2}G_Fn(r)}{m_n}\tfrac{1-Y_e}{2})
\ket{\nu_x}\bra{\nu_x}\\
+&\,\tfrac{\delta m^2}{2E_\nu} \sin{2\theta}\, 
\left(\ket{\nu_e}\bra{\nu_x}+\ket{\nu_x}\bra{\nu_e}\right)
\end{split}
\end{equation}
for neutrinos and 
\begin{equation} 
\label{Hnubarnubar}
\begin{split}
H_{\bar\nu \leftrightarrow \bar\nu}(r)
=&\qty(\tfrac{\delta m^2}{2E_\nu}\cos{2\theta}+\tfrac{\sqrt{2}G_Fn(r)}{m_n}\tfrac{1-Y_e}{2})
\ket{\bar\nu_x}\bra{\bar\nu_x} \\
+&\qty(-\tfrac{\delta m^2}{2E_\nu}\cos{2\theta}-\tfrac{\sqrt{2}G_Fn(r)}{m_n}\tfrac{3Y_e-1}{2})
\ket{\bar\nu_e}\bra{\bar\nu_e}\\
+&\,\tfrac{\delta m^2}{2E_\nu} \sin{2\theta}\, 
\left(\ket{\bar\nu_e}\bra{\bar\nu_x}+\ket{\bar\nu_x}\bra{\bar\nu_e}\right)
\end{split}
\end{equation}
for antineutrinos. Here, the terms involving the angle $\theta$
represent the vacuum mixing of flavor eigenstates. $E_\nu$ denotes the neutrino
energy, and $\delta m^2$ denotes the squared difference of the two neutrino
masses. In NH we have $\delta m^2>0$ whereas in IH we have $\delta m^2<0$. For
the $13$ mixing parameters adopted here, we have $\sin 2\theta=0.29$ and
\begin{equation}
\label{vacuum value}
\frac{|\delta m^2|}{2E_\nu}=\frac{6.4\times 10^{-11}\mbox{eV}}{E_\nu/10 \mbox{
MeV}}, 
\end{equation}
which is the energy separation between mass eigenstates in vacuum. 

\begin{figure}
\centering
\includegraphics[width=.45\textwidth]{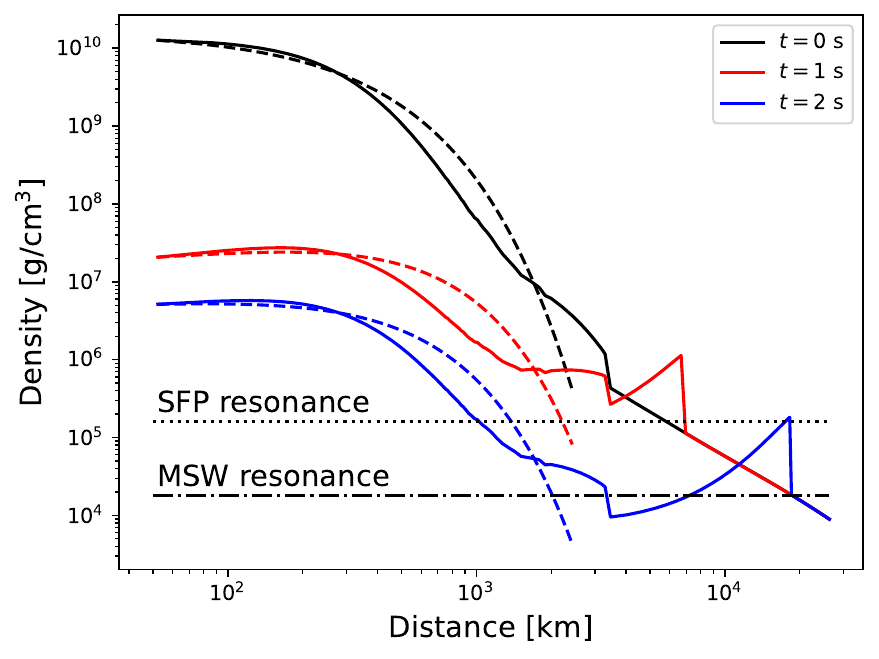}
\caption{The density versus distance inside the supernova. The solid black line
shows the density of the $6M_\odot$ helium core presupernova model taken from
Ref. \cite{1987ESOC...26..325N}. The solid red and blue lines show the
parametric shock wave superimposed on the presupernova distribution as described
in Ref. \cite{Fogli:2003dw} at $t=1$ s and $t=2$ s, respectively. The dashed
lines in corresponding colors show the fitted density distributions described by
Eqs. (\ref{density profile}) and (\ref{fits}). For later post-bounce times, the
results are close to the blue lines and not shown here. The horizontal lines
show the SFP and MSW resonance densities for a $1$ MeV neutrino.} 
\label{fig:baryonProfileShock}
\end{figure}

In Eqs. (\ref{Hnunu}) and (\ref{Hnubarnubar}), the environment is characterized
by its mass density $n(r)$ and its electron fraction $Y_e$. The electron fraction is
defined as the ratio of electron and baryon number densities. $m_n$ denotes the
average baryon mass. In our calculations we assume slightly neutron rich
conditions by taking $Y_e=0.45$. The density profile that we use is based on the
$6M_{\odot}$ helium core presupernova model of Ref. \cite{1987ESOC...26..325N}.
We adopt this as the density distribution at the shock bounce at $t=0$. At later
times, the shock wave modifies the density profile. We mimic this by
parametrically changing the $t=0$ matter density as described in Ref.
\cite{Fogli:2003dw} for up to post-bounce time $t=5$ s. We do this by using same
shock wave speed and contact discontinuity parameters as in Ref.
\cite{Fogli:2003dw}. The resulting density distributions are shown in Fig.
\ref{fig:baryonProfileShock} with solid lines. These are the density
distributions that we later use for our realistic calculations in Section VII. 
But, before that we run some illustrative simulations by fitting these density
distributions to the functional form
\begin{equation}
\label{density profile}
n(r)=n_0 e^{-r/r_{\mbox{\footnotesize{mat}}}}.
\end{equation}
We find that, at all post-bounce times and for the first few thousand
kilometers, Eq. (\ref{density profile}) fits the density distributions
reasonably well with $r_{\mbox{\footnotesize{mat}}}=200$ km and with different
central densities at different post-bounce times. At $t=0$, the central density
is $n_0=1.0\times 10^{10} \mbox{ g/cm}^3$. At later times we have
\begin{equation}
\label{fits}
n_0=
\begin{cases}
1.8\times 10^7 \mbox{g/cm}^3 & t=1 \mbox{ s}, \\
4.4\times 10^6 \mbox{g/cm}^3 & t=2 \mbox{ s}, \\
2.3\times 10^6 \mbox{g/cm}^3 & t=3 \mbox{ s}, \\
1.5\times 10^6 \mbox{g/cm}^3 & t=4 \mbox{ s}, \\
1.0\times 10^6 \mbox{g/cm}^3 & t=5 \mbox{ s}.
\end{cases}
\end{equation}
The fitted densities are also shown in Fig. \ref{fig:baryonProfileShock} with
dashed lines. The practical reason for using the fitted distributions in
illustrative simulations is the necessity of running a large number of
simulations in order to discuss various features of the phase effect. But
another rationale is the fact that the SFP phase effect actually takes place in
the density scale: the results mostly depend on the value of the magnetic field
at a particular density rather than how the density and magnetic field are
spread over the physical space. In other words, it is the interplay between the
matter profile and the magnetic field profile that matters. It is in that sense
that Eqs. (\ref{B value}) and (\ref{density profile}) serve as a basis for
illustration. For example, Fig. \ref{fig:baryonProfileShock} clearly shows
that, in the realistic case, the density decreases more slowly and the
resonances take place in outer regions where the magnetic field is weaker.
For this reason, a sizable SFP phase effect requires either a larger
magnetic field or a larger neutrino magnetic moment in the realistic case than
it is in the illustrative case. A shortcoming associated with using the
exponential fits is that this approach reduces the post-bounce supernova
dynamics to one variable, which is the central density given in Eq.
(\ref{fits}). In particular, with the fitted densities neutrinos pass through
MSW resonance only once whereas with actual densities some neutrinos pass it
three times. However, as we discuss further in the Conclusions, additional resonances can
only serve to increase the uncertainties associated with SFP phase effect. 

On the surface of the proto-neutron star, the energy separation between flavor
eigenstates created by the interactions is of the order of 
\begin{equation}
\label{numerical densities}
7.8 \times 10^{-9} \mbox{ eV} \leq\frac{\sqrt{2}G_Fn_0}{m_n}(1-2Y_e)
\leq 1.4\times10^{-7} \mbox{ eV},
\end{equation}
where the minimum value corresponds to $t=1$ s and the maximum value corresponds to $t=5$ s. Comparing
Eq. (\ref{numerical densities}) with Eq. (\ref{muB value}) one might think that
the importance of the SFP phase effects increases at later post-bounce times.
However, as we discuss below, the situation less straightforward. 

\section{The Resonances}

Since neutrinos of all flavors are emitted from the neutrinosphere, it is better
to work with a density operator rather than individual initial states. We denote
by $\hat{\rho}(E_\nu,r)$ the normalized density operator which describes all
neutrinos and antineutrinos with energy $E_\nu$ which are at a distance $r$ from
the center of the supernova. The energy dependence of $\hat\rho(E_\nu,r)$ comes
in part from vacuum oscillations as shown in Eq. (\ref{vacuum value}), and in
part from the energy dependence of emission from the neutrinosphere. For
brevity, we suppress the energy dependence of the density operator unless it is
necessary for discussion. But whenever we write $\hat\rho(r)$, a specific
neutrino energy is always implied. 

We assume that all neutrinos represented by the density operator $\hat\rho(r)$
are emitted from the surface of the neutrinosphere at the same time and traveled
outward with the speed of light to reach $r$. This is a simplification
because in reality every point on the surface of the neutrinosphere emits
neutrinos uniformly in every direction pointing outward. Even in a spherically
symmetric supernova, which we assume to be the case, neutrinos would travel by
slightly different distances and experience slightly different conditions 
depending on their direction of emission. Ignoring this dependence is called the single
angle approximation\footnote{The geometry of this approximation is studied in
detail in Ref. \cite{Duan:2006an}. The focus of this reference is the
neutrino-neutrino interactions, but its geometrical treatment of neutrinos
emitted in all directions from a spherical source can be generally applied.} and
it effectively reduces the problem to the one dimensional evolution equation 
\begin{equation}
\label{eqn:EoM}
i\dv{}{r}\hat{\rho}(r) = \comm{H(r)}{\hat{\rho}(r)}. 
\end{equation}
Here the flavor evolution is described in terms of the distance $r$ because
neutrinos essentially travel with the speed of light. Most of the non-trivial
flavor evolution takes place by the time neutrinos reach the low density outer
layers of the supernova, which takes only a fraction of a second. This is much
shorter than the time scale with which the supernova background changes. For
this reason, one can take a snapshot of the supernova at the post-bounce time
$t$, put this information into the Hamiltonian $H(r)$ in Eq. (\ref{eqn:EoM})
and solve it to find the flavor evolution of the neutrinos that are emitted at
time $t$. In this sense, Eq. (\ref{eqn:EoM}) depends on the post-bounce time
$t$. This dependence is not explicitly shown in our equations, but we solve Eq.
(\ref{eqn:EoM}) for the post-bounce times from $t=1$ s to $t=5$ s and clearly
label the corresponding results. 

The total Hamiltonian $H(r)$ includes the vacuum oscillations of neutrinos, their
interactions with the matter background, and the effect of the magnetic field.
It can be written as 
\begin{equation}
\label{distant decomposition}
H(r) = H_{\nu \leftrightarrow \nu}(r) + H_{\bar\nu \leftrightarrow \bar\nu}(r) +
H_{\mu}(r)
\end{equation}
with the individual terms given by Eqs. (\ref{Hmu}), (\ref{Hnunu}) and
(\ref{Hnubarnubar}). The evolution described by Eqs. (\ref{eqn:EoM}) and
(\ref{distant decomposition}) is not difficult to solve numerically. However,
much insight can be gained by analytically examining its evolution under
adiabatic conditions.  In general, adiabaticity refers to a situation where the
external conditions affecting a system change slowly in comparison to the system
itself. For the problem at hand, this means that the distance scales over which
the matter density and the magnetic field change should be much longer than the
distance scale over which the neutrino oscillates. Adiabatic evolution can be
described in a simple way in terms of the energy eigenstates of the Hamiltonian:
An initial state which is nearly an energy eigenstate evolves approximately into
the same eigenstate at later times. We can quantify this by denoting the local
energy eigenvalues of the total Hamiltonian in Eq. (\ref{distant decomposition})
by $E_i(r)$ and the corresponding energy eigenstates by $|r_i\rangle$. In other
words, at every point $r$, we have
\begin{equation}
\label{local eigenstates}
\hat H (r)= \sum_{i=1}^4 E_i(r)\ket{r_i}\bra{r_i}.
\end{equation}
We order the eigenvalues such that $E_1$ is the largest and $E_4$ is the
smallest. The adiabaticity condition can be expressed as
\begin{equation}
\label{adiabaticity condition}
\abs{\mel{r_i}{\dv{H(r)}{r}}{r_j}}\ll|E_i(r)-E_j(r)|.
\end{equation}
As long as the adiabaticity requirement is met, the evolution is determined
solely by the energy spectrum of the Hamiltonian. The adiabatic approximation
tells us that if the initial state of a neutrino on the neutrinosphere ($r=R$)
is an energy eigenstate $|R_i\rangle$, then it evolves into
\begin{equation}
\label{adiabatic evolution}
|R_i\rangle \rightarrow e^{-i\int_R^r E_i(r)dr} |r_i\rangle
\end{equation}
at a later time. 
\begin{figure}
\centering
\includegraphics[width=0.82\columnwidth]{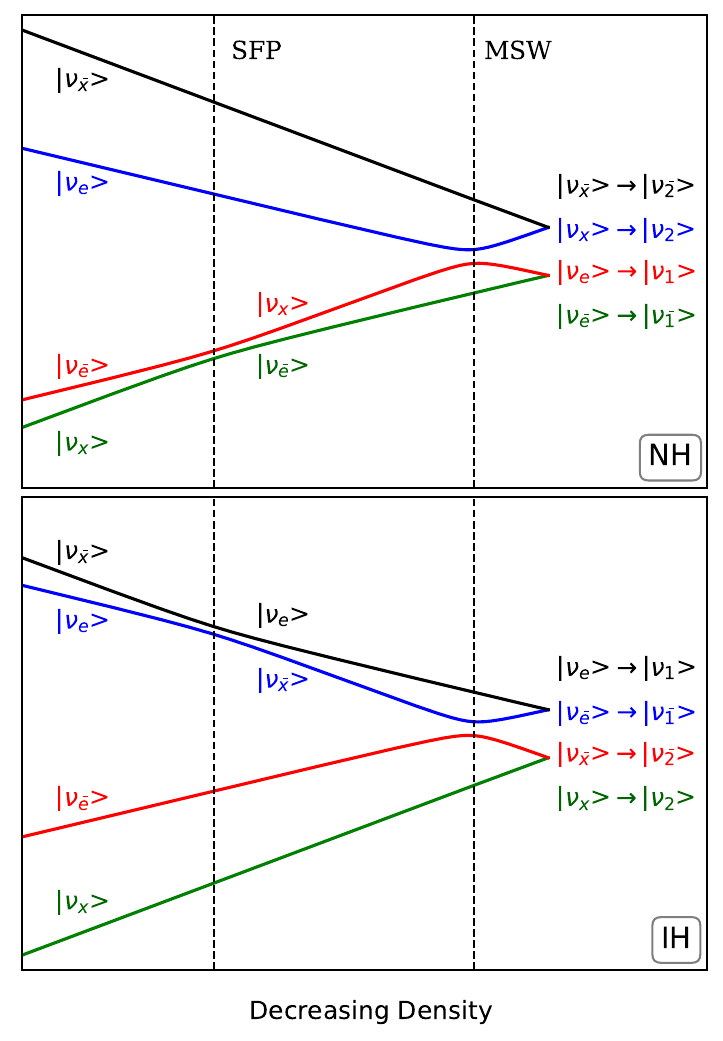}
\caption{The energy eigenvalues of the total Hamiltonian as functions of the
logarithm of the density, which decreases from left to right. The figure is
independent of the density distribution model. Dominant flavor contents of the
corresponding eigenstates are indicated next to the lines. On the right, we show
the flavor contents of the eigenstates right after the MSW resonance and to
which mass eigenstates they adiabatically evolve in vacuum.}
\label{eigenvalues}
\end{figure}

Fig. \ref{eigenvalues} shows the eigenvalues of the Hamiltonian as functions of
the logarithm of the density. Since the eigenvalues are plotted against
density, the actual density distribution is irrelevant. The two points where
the eigenvalues approach to each other pairwise are SFP and MSW resonance
points.  These are the points at which the adiabaticity condition comes closest
to being violated.  If an adiabaticity violation occurs, one speaks of a
partially adiabatic resonance. In partially adiabatic case, even if the initial
state is approximately an energy eigenstate, it evolves into a superposition of
the two approaching energy eigenstates as described by the Landau-Zener jumping
probability \cite{1932PhyZS...2...46L, 1932RSPSA.137..696Z,
1981PhRvA..23.3107R} to be discussed below.  As the density decreases from the
center, the first resonance occurs between $E_4$ and $E_3$ for NH, and between
$E_1$ and $E_2$ for IH. On the other hand, near the central regions the density
is large enough to force each energy eigenstate to significantly project on a
particular flavor eigenstate. The dominant flavor content of each energy
eigenstate is indicated in Fig. \ref{eigenvalues}. In particular, noting that
$3Y_e-1<1-Y_e$ for the neutron rich conditions ($Y_e<1/2$), and substituting
the numerical values given in Eqs. (\ref{muB value}), (\ref{vacuum value}), and
(\ref{numerical densities}), into the Hamiltonians (\ref{Hnunu}) and
(\ref{Hnubarnubar}), we find
\begin{equation}
\label{energy eigenkets at R}
\left(\ket{R_1},\ket{R_2},\ket{R_3},\ket{R_4}\right)\approx
\left(\ket{\bar\nu_x},\ket{\nu_e},\ket{\bar\nu_e},\ket{\nu_x}\right) 
\end{equation}
in both hierarchies on the neutrinosphere. In that sense, one can loosely say
that the SFP resonance occurs between $\bar\nu_e-\nu_x$ in NH, and between
$\nu_e-\bar\nu_x$ in IH. Indeed, one can re-write the Hamiltonian in Eq.
(\ref{distant decomposition}) in the following decomposition: 
\begin{equation}
\label{early decomposition}
H(r) = H_{e \leftrightarrow \bar x}(r) + H_{x \leftrightarrow \bar e}(r) + H_{\theta}.
\end{equation}
Here, $H_{e \leftrightarrow \bar x}(r)$ and $H_{x \leftrightarrow \bar e}(r)$
are the parts of the Hamiltonian in Eq. (\ref{distant decomposition}) which live in
the orthogonal $\nu_e-\bar\nu_x$ and $\bar\nu_e-\nu_x$ subspaces, respectively.
They are given by
\begin{align}
\label{H_exbar}
H_{e \leftrightarrow \bar x}(r) 
=-&\qty(\tfrac{\delta m^2}{4E}\cos{2\theta}\!-\!\tfrac{\sqrt{2}G_F n(r)}{m_n}\tfrac{3Y_e-1}{2})
\ket{\nu_e}\bra{\nu_e}\nonumber\\
+&\qty(\tfrac{\delta m^2}{4E}\cos{2\theta}\!+\!\tfrac{\sqrt{2}G_F n(r)}{m_n}\tfrac{1-Y_e}{2})
\ket{\bar\nu_x}\bra{\bar\nu_x}\nonumber\\
+& \, \mu B \, \left(\ket{\nu_e}\bra{\bar\nu_x} + \ket{\bar\nu_x}\bra{\nu_e}\right)
\end{align}
and
\begin{align}
\label{H_xebar}
H_{\bar e \leftrightarrow x}(r) 
=-&\qty(\tfrac{\delta m^2}{4E} \cos{2\theta}\!+\!\tfrac{\sqrt{2}G_F n(r)}{m_n}\tfrac{3Y_e-1}{2})
\ket{\bar\nu_e}\bra{\bar\nu_e}\nonumber\\
+&\qty(\tfrac{\delta m^2}{4E} \cos{2\theta}\!-\!\tfrac{\sqrt{2}G_F n(r)}{m_n}\tfrac{1-Y_e}{2})
\ket{\nu_x}\bra{\nu_x}\nonumber\\
-& \, \mu B \, \left(\ket{\nu_x}\bra{\bar\nu_e}+\ket{\bar\nu_e}\bra{\nu_x}\right).
\end{align}
These two Hamiltonians describe flavor transition in two orthogonal channels.
The term $H_\theta$ in Eq. (\ref{early decomposition}) is given by 
\begin{equation*}
\label{Htheta}
H_{\theta}=
\tfrac{\delta m^2}{2E_\nu} \sin{2\theta} \qty(\ket{\nu_e}\bra{\nu_x}+\ket{\nu_x}\bra{\nu_e}
+\ket{\bar\nu_e}\bra{\bar\nu_x}+\ket{\bar\nu_x}\bra{\bar\nu_e})
\end{equation*}
and it couples these two orthogonal channels. Since $H_{\theta}$ is proportional
to the small term $\tfrac{\delta m^2}{2E_\nu}\sin\theta$, its effect can be
ignored near the neutrinosphere. In this case, the spin-flavor evolution
proceeds through the decoupled $\nu_e-\bar\nu_x$ and $\bar\nu_e-\nu_x$ channels.
This is what we see on the high density part of Fig. \ref{eigenvalues} where the
large energy separation between the upper and lower pairs of eigenvalues forbids
the transition between them. In this decoupling approximation, SFP resonance
occurs when the diagonal elements of the Hamiltonians in Eqs. (\ref{H_exbar}) or
(\ref{H_xebar}) become equal. This happens when \cite{Lim:1987tk,
Akhmedov:1988uk} 
\begin{equation}
\label{sf resonance}
\frac{\delta m^2}{2E_\nu} \cos 2\theta = \pm
\frac{\sqrt{2}G_Fn(r)}{m_n}(1-2Y_e),
\end{equation}
where $-$ sign is for $H_{e \leftrightarrow \bar x}(r)$ and $+$ sign is for
$H_{\bar e \leftrightarrow x}(r)$. Clearly, for $Y_e<0.5$ adopted in this paper,
the condition in Eq. (\ref{sf resonance}) can hold only for the former in NH,
and only for the latter in IH. In the rest of this paper, we focus on NH. When
the effect of the $H_{\theta}$ is included the location of the SFP resonance
shifts as discussed in Ref. \cite{Friedland:2005xh}. But for the small mixing
angle that we use, this is inconsequential. 
\begin{figure}
\centering
\includegraphics[width=0.85\columnwidth]{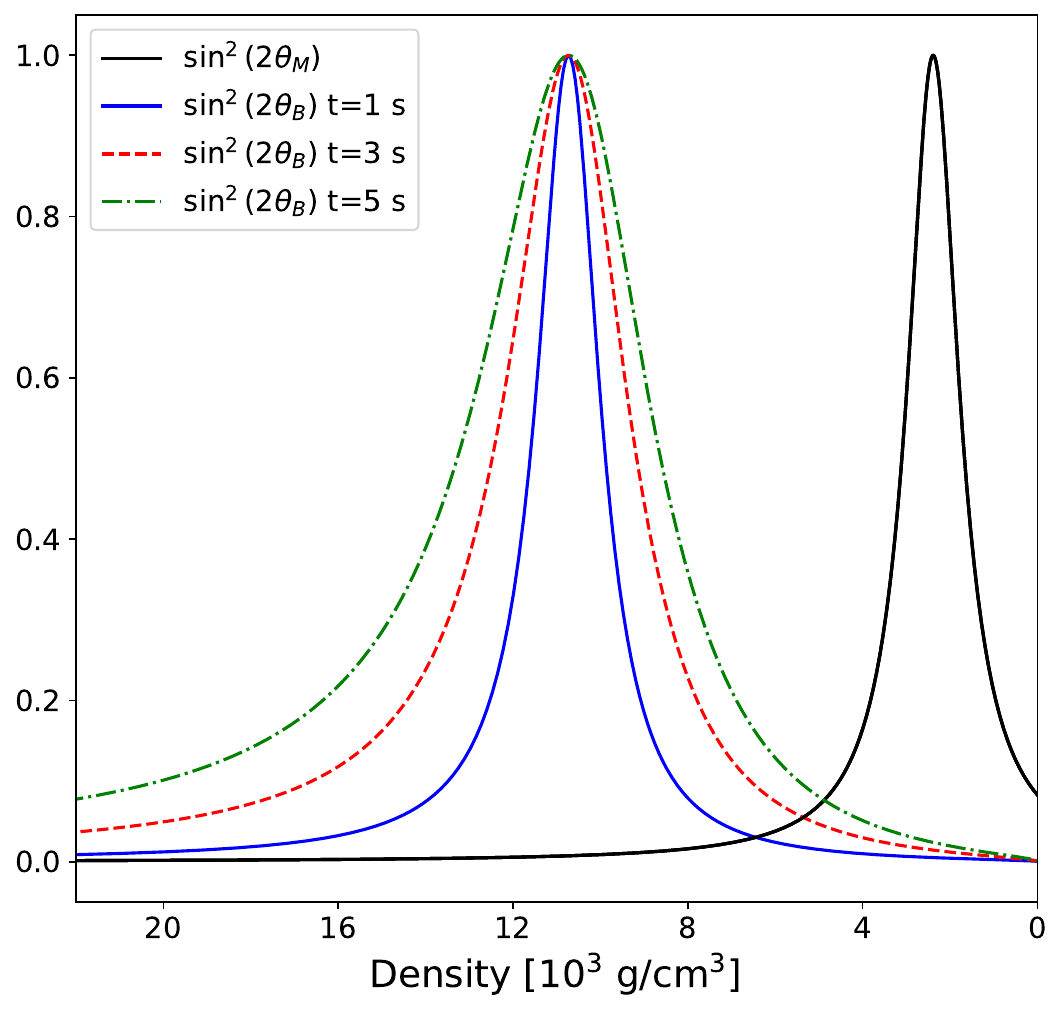}
\caption{Resonance widths as represented by effective mixing angles at different
post-bounce times. The density decreases from left to right. The solid black
line shows $\sin^2\!2\theta_M(r)$ corresponding to the MSW resonance, whose
universality means that its position and width is the same in density scale at
all post-bounce times. The solid blue, dashed red, and dash-doted green lines
show $\sin^2\!2\theta_B(r)$ for SFP resonance at post-bounce times $t=1,3,5$ s,
respectively. While SFP resonance occurs at a fixed density, its physical
location moves to inner regions at later post-bounce times where the magnetic
field is stronger. For this reason it becomes wider with time. The figure is for
a $15$ MeV neutrino with exponentially fitted density distributions. The value
of $\mu B_0$ given by Eq. (\ref{muB value}) is used. For smaller $\mu B_0$, the
SFP resonance is narrower.}
\label{resonance widths}
\end{figure}

In the decoupling limit, Eq. (\ref{H_xebar}) tells us that the lower two
eigenstates which undergo SFP resonance are combinations of $\ket{\nu_x}$ and
$\ket{\bar\nu_e}$. This can be expressed in terms of an effective mixing angle: 
\begin{equation}
\begin{split}
\label{local eigenkets SFP}
\ket{r_3}&=\cos\theta_B(r) \ket{\nu_x} + \sin\theta_B(r) \ket{\bar\nu_e}, \\
\ket{r_4}&=-\sin\theta_B(r) \ket{\nu_x}+ \cos\theta_B(r) \ket{\bar\nu_e}.
\end{split}
\end{equation}
Here the effective mixing angle $\theta_B(r)$ is defined in
the range $0\leq \theta_B(r) \leq \pi/2$ with 
\begin{equation}
\label{theta SFP}
\tan{2\theta_B(r)}=\frac{2\mu B(r)}{\frac{\delta
m^2}{2E}\cos\theta-\tfrac{\sqrt{2}G_Fn(r)}{m_n}(1-2Y_e)}. 
\end{equation}
Well above the resonance density, $\theta_B(r)\simeq\pi/2$ so that the
eigenstates are $\ket{r_3}\approx\ket{\bar\nu_e}$ and
$\ket{r_4}\approx-\ket{\nu_x}$ as expected from Fig. \ref{eigenvalues}. Well
below the resonance density $\theta_B(r)\simeq 0$ in which case the flavor
contents of eigenstates are switched as is also shown in Fig. \ref{eigenvalues}.
At the resonance density where $\theta_B=\pi/4$, the energy eigenstates are
maximal mixtures of flavor eigenstates. In that sense 
\begin{equation}
\label{sin 2theta}
\sin^2 \! 2\theta_B(r)=4\sin^2 \! \theta_B(r) \cos^2 \! \theta_B(r) 
\end{equation}
can be used a measure of the width of the resonance because it will be different
from zero as long as energy eigenstates are mixtures of flavor eigenstates. This
quantity is plotted in Fig. \ref{resonance widths} against density at different
post-bounce times. The solid blue line is for $t=1\mbox{ s}$, the red dashed
line is for $t=3\mbox{ s}$, and green dashed-doted line for $t=5\mbox{ s}$. The
points at which $\sin^2\!2\theta_B=1$ (i.e., the resonance points) coincide
because the plot is in density scale and SFP resonance occurs at a specific
density. But the physical location of the SFP resonance moves closer to the
center of the supernova at later post-bounce times as the overall density drops.
This can also be seen in Fig. \ref{fig:baryonProfileShock} where the horizontal
line representing the SFP resonance crosses the density distributions at
increasingly smaller radii. While this is true for both the realistic and the
fitted density distributions, the latter is used in Fig. \ref{resonance widths}.
An important observation is that the SFP resonance region, i.e. the region for
which $\sin^2\!2\theta_B(r)$ is significantly different from zero, becomes
increasingly wider in density scale at later post-bounce times. It happens
because as the resonance moves inward it occurs in a region where the magnetic
field is stronger. According to Eq. (\ref{theta SFP}) a stronger magnetic field
at a given density means a larger $\theta_B(r)$. The widening of SFP resonance
with post-bounce time will be important in what follows.
However, there is no similar widening effect for the MSW resonance. 

MSW resonance occurs in the outer regions where the magnetic field is weaker.
For this reason, $H_\mu(r)$ can be ignored in the decomposition given by Eq.
(\ref{distant decomposition}) so that the neutrino and antineutrino
oscillations decouple. In this case, Fig.  \ref{eigenvalues} tells us that the
dynamics of $\ket{r_1}-\ket{r_4}$ decouple from the dynamics of
$\ket{r_2}-\ket{r_3}$ in both hierarchies. With this approximation, MSW
resonance occurs when the diagonal elements of $H_{\nu \leftrightarrow \nu}(r)$
or $H_{\bar\nu \leftrightarrow \bar\nu}(r)$ become equal, i.e. when 
\begin{equation}
\label{msw resonance}
\frac{\delta m^2}{2E_\nu} \cos 2\theta = \pm
\frac{\sqrt{2}G_Fn(r)}{m_n}Y_e.
\end{equation}
Here the $+$ sign is for neutrinos and $-$ signs is for antineutrinos.
Resonance condition in Eq. (\ref{msw resonance}) holds only for neutrinos
in NH, and only for antineutrinos in IH. Since we focus on NH, 
the resonating eigenstates can be written as 
\begin{equation}
\begin{split}
\label{local eigenkets MSW}
\ket{r_2}&=\cos\theta_M(r) \ket{\nu_e} + \sin\theta_M(r) \ket{\nu_x}, \\
\ket{r_3}&=- \sin\theta_M(r) \ket{\nu_e} + \cos\theta_M(r) \ket{\nu_x} 
\end{split}
\end{equation}
in terms of an effective matter mixing angle $\theta_M(r)$ defined in the range
$0\leq \theta_M(r)\leq \pi/2$ and given by 
\begin{equation}
\label{theta MSW}
\tan{2\theta_M(r)}=\frac{\tfrac{\delta m^2}{2E_\nu}\sin2\theta}{\tfrac{\delta
m^2}{2E_\nu}\cos2\theta-\frac{\sqrt{2}G_Fn(r)}{m_n}Y_e}.
\end{equation}
The quantity $\sin^2\!2\theta_M(r)$ is similarly a measure of the resonance
width and is plotted in Fig. \ref{resonance widths}. Like the SFP resonance,
the MSW resonance occurs at the same position in density scale but physically
moves inward with time. However, unlike the SFP resonance its width is fixed in
density scale. This is because at a given density the value of $\theta_M(r)$
depends only on the vacuum mixing parameters. This is referred to as the
universality. For a detailed discussion, see, for example, Ref.
\cite{Smirnov:2003da}. The MSW resonance is universal in the sense that its
width depends only on the vacuum mixing parameters at the density scale, not on
how this density is spread in physical space.  In that sense SFP resonance is
not universal because its width depends on how the magnetic field changes with
respect to the density.

The above description relies on the decoupling of the $\ket{r_1}-\ket{r_2}$ and
$\ket{r_3}-\ket{r_4}$ dynamics in the SFP resonance region, and the decoupling
of $\ket{r_1}-\ket{r_4}$ and $\ket{r_2}-\ket{r_3}$ dynamics in the MSW resonance
region. These two approximations are not independent. They are both true when
SFP and MSW resonances are well separated, and they both fail when these
resonances are not well separated. The eigenstate $\ket{r_3}$ which enters both
resonances is the culprit here. We write $\ket{r_3}$ as a combination of
$\ket{\nu_x}$ and $\ket{\bar\nu_e}$ around the SFP resonance in Eq. (\ref{local
eigenkets SFP}), and as a combination of $\ket{\nu_x}$ and $\ket{\nu_e}$ around
the MSW resonance in Eq. (\ref{local eigenkets MSW}).  Obviously both equations
can be true only if $\sin^2\!2\theta_B(r)$ drops nearly to zero by the time
$\sin^2\!2\theta_M(r)$ starts to become different from zero. Otherwise
$\ket{r_3}$ has to be combination of $\ket{\nu_x}$, $\ket{\bar\nu_e}$, and
$\ket{\nu_e}$. Therefore there is only one decoupling approximation, and it is
the decoupling of the SFP and MSW resonances as quantified by their widths.
Fig. \ref{resonance widths} tells us that in general the decoupling
approximation can be valid at early post-bounce times, but likely to fail
at late post-bounce times. The degree to which it fails depends on how the
magnetic field changes with respect to matter density, and how large the
neutrino magnetic moment is because SFP resonance width is directly controlled
by the value of $\mu B(r)$ around the resonance region.

\section{Evolution and Decoherence}

In what follows, we first consider electron antineutrino survival
probabilities. This serves to illustrate the appearance of SFP phase effect and
discuss its relative importance under different settings. We work with electron
antineutrinos because they undergo SFP resonance in NH. To do this, we start
with a pure ensemble of electron antineutrinos represented by the initial
density operator 
\begin{equation}
\label{box}
\hat \rho (E_\nu,R)= \ket{\bar\nu_e}\bra{\bar\nu_e}
\end{equation}
and then calculate its evolution. If the density operator evolves into $\hat\rho(E_\nu,r)$ at
$r$, then the survival probability is given by 
\begin{equation}
\label{general survival probability}
P_{\bar\nu_e\to\bar\nu_e}(E_\nu,r) = \mel{\bar\nu_e}{\hat\rho(E_\nu,r)}{\bar\nu_e}.
\end{equation}
We calculate this both numerically by solving the evolution equation given in
Eq. (\ref{eqn:EoM}), and  analytically by using the decoupling approximation
together with the Landau-Zener jumping probabilities. Note that here we
temporarily reintroduce the neutrino energy $E_\nu$ into our notation to
emphasize the energy dependence of the survival probability. 

After that, we consider a mixed ensemble of neutrinos as appropriate for the
cooling period of the proto-neutron star. In this case, the initial
density operator is in the diagonal form 
\begin{equation}
\label{initial R in flavor basis}
\hat \rho (E_\nu,R)= \sum_{\alpha=e,\bar e, x, \bar x}
\rho_{\alpha\alpha}(E_\nu,R)\ket{\nu_\alpha}\bra{\nu_\alpha},
\end{equation}
where $\rho_{\alpha\alpha}(E_\nu,R)$ is the $\nu_\alpha$ energy spectrum emitted
from the neutrinosphere in arbitrary units. Here, $\alpha= e,\bar e, x, \bar x$
and we use $\nu_{\bar \alpha}$ to mean $\bar\nu_\alpha$. We again calculate the
corresponding density operator $\hat\rho(E_\nu,r)$ at $r$ both numerically and
analytically as described above. Corresponding $\nu_\alpha$ energy spectrum at $r$
is given by 
\begin{equation}
\label{distributions}
\rho_{\alpha\alpha}(E_\nu,r)=\mel{\nu_\alpha}{\hat\rho(E_\nu,r)}{\nu_\alpha}. 
\end{equation}

In both cases (for the pure ensemble of electron antineutrinos and for the
mixed ensemble) analytical calculations start by expressing the initial density
operator in energy eigenbasis at $R$ as 
\begin{equation}
\label{initial R in eigenbasis}
\hat\rho(R) = \sum_{i,j=1}^4 \rho_{ij}(R) \ket{R_i} \bra{R_i}
\end{equation}
with $\rho_{ij}(R)=\mel{R_i}{\hat\rho(R)}{R_j}$ denoting the corresponding
components. Here, we again drop the energy dependence from our notation. We
emphasize that although Latin indicies are typically used to refer mass basis
in the literature, we use them to refer to energy eigenbasis in this paper.

Let us first assume that the adiabaticity condition holds through both
resonances. We consider the partially adiabatic evolution in the next two
sections. In the adiabatic case, the evolution is completely determined by Eq.
(\ref{adiabatic evolution}). The density operator evolves into 
\begin{equation}
\label{later rho in eigenbasis}
\hat\rho(r)= \sum_{i,j=1}^4 e^{-i\int_R^r(E_i-E_j)dr} \rho_{ij}(R) \ket{r_i} \bra{r_i}
\end{equation}
at a later $r$. As the neutrinos reach to the surface of the supernova, the
Hamiltonian is reduced to the vacuum term alone and the energy eigenstates
reduce to the mass eigenstates. Which energy eigenstate reduces to which mass
eigenstate depends on the hierarchy. We have 
\begin{equation}
\label{eigenstates in vacuum}
\begin{split}
\ket{r_1}\!,\ket{r_2}\!,\ket{r_3}\!,\ket{r_4} \! \xrightarrow{n(\!r\!)\to 0} \!
\ket{\bar\nu_2}\!,\ket{\nu_2}\!,\ket{\nu_1}\!,\ket{\bar\nu_1} 
& \!\mbox{ in NH,}\\ 
\ket{r_1}\!,\ket{r_2}\!,\ket{r_3}\!,\ket{r_4} \! \xrightarrow{n(\!r\!)\to 0} \!
\ket{\nu_1}\!,\ket{\bar\nu_1}\!,\ket{\bar\nu_2}\!,\ket{\nu_2} 
& \!\mbox{ in IH.}\\ 
\end{split}
\end{equation}
These are also indicated in Fig. \ref{eigenvalues}.

After that, neutrinos travel a long distance to Earth and during this time they
decohere. Decoherence happens because neutrino mass eigenstates travel with
different speeds in vacuum and a gap opens up between them over long distances.
When neutrino mass eigenstates do not overlap in physical space, they cannot
interfere and the flavor oscillations stop \cite{GIUNTI199287}. The standard
mathematical formulation of neutrino oscillations, which we also use here, is
based on the assumption that neutrinos are plain waves with infinite wavepackage
size. The formulation of the decoherence requires the finite size of the
wavepackage to be taken into account. In this case, the off diagonal elements of
the density matrix in mass basis pick up an exponential term
$e^{-(r/r_{\mbox{\tiny coh}})^2}$ \cite{Hansen:2016klk}. The coherence length
$r_{\mbox{\tiny coh}}$ depends on the wave package size, which in turn depends
on the circumstances of the neutrino's creation. For a physically intuitive
discussion, see Section 8 of Ref. \cite{Giunti}. For supernova neutrinos the
coherence length can be estimated to be of the order of $10^6$ km \cite{Giunti,
Nussinov:1976uw}. For this reason, plane wave formulation is a good
approximation inside the star. But even for a galactic supernova, neutrinos have
to travel several kilo parsecs (about $10^{17}$ km) to reach Earth. Therefore, one
can take $r\to\infty$ limit where $e^{-(r/r_{\mbox{\tiny coh}})^2}\to 0$.
Therefore the decoherence of neutrinos over long distances can be implemented in
practice by removing the off-diagonal elements of the density operator in the
mass basis. As a result, in the fully adiabatic case the neutrinos arriving
Earth is described by 
\begin{equation}
\label{adiabatic rho earth}
\hat\rho(\infty) = \sum_{i=1}^4 \rho_{ii}(R) \ket{r_i} \bra{r_i}
\end{equation}
with $\ket{r_i}$ given by Eq. (\ref{eigenstates in vacuum}) according to mass
hierarchy. 
In particular, for the initial pure ensemble of electron antineutinos given in Eq. (\ref{box}), one finds
\begin{equation}
\label{rho ebar -> infinity adiabatic}
\hat\rho(\infty)=\sin^2\!\theta_B(R)\ket{\nu_1}\bra{\nu_1}+\cos^2\!\theta_B(R)\ket{\bar\nu_1}\bra{\bar\nu_1}. 
\end{equation}
Here we used Eq. (\ref{local eigenkets SFP}) to calculate $\rho_{ii}(R)$ in
terms of the effective mixing angle $\theta_B(R)$ on the surface of the
neutrinosphere. Using the fact that $\bra{\nu_1}\ket{\bar\nu_e}=0$ and
$\bra{\bar\nu_1}\ket{\bar\nu_e}=\cos\theta$, the survival probability of an
initial $\bar\nu_e$ is found from Eq. (\ref{general survival probability}) as 
\begin{equation} 
\label{Pebarebar} 
P_{\bar\nu_e\to\bar\nu_e}(\infty)
=\left(\frac{1}{2}+\frac{1}{2}\cos2\theta_B(R)\right)\cos^2\!\theta,
\end{equation}
which is the standard result. See Ref. \cite{1992PhR...211..167P}, for
example. 

\section{Zero mixing angle}

\subsection{General treatment}

Here and in the next section, we consider the partially adiabatic evolution of
neutrinos. In this section we take $\theta=0$ which removes the MSW resonance
from the picture and allows us to focus on the phase effects between the
production point and the partially adiabatic SFP resonance. $\theta\neq 0$ case
is discussed in the next section where additional phase effects from the MSW
resonance also enter into the picture. 
\begin{figure*}[hbt!]
\begin{center}
\includegraphics[width=0.97\textwidth]{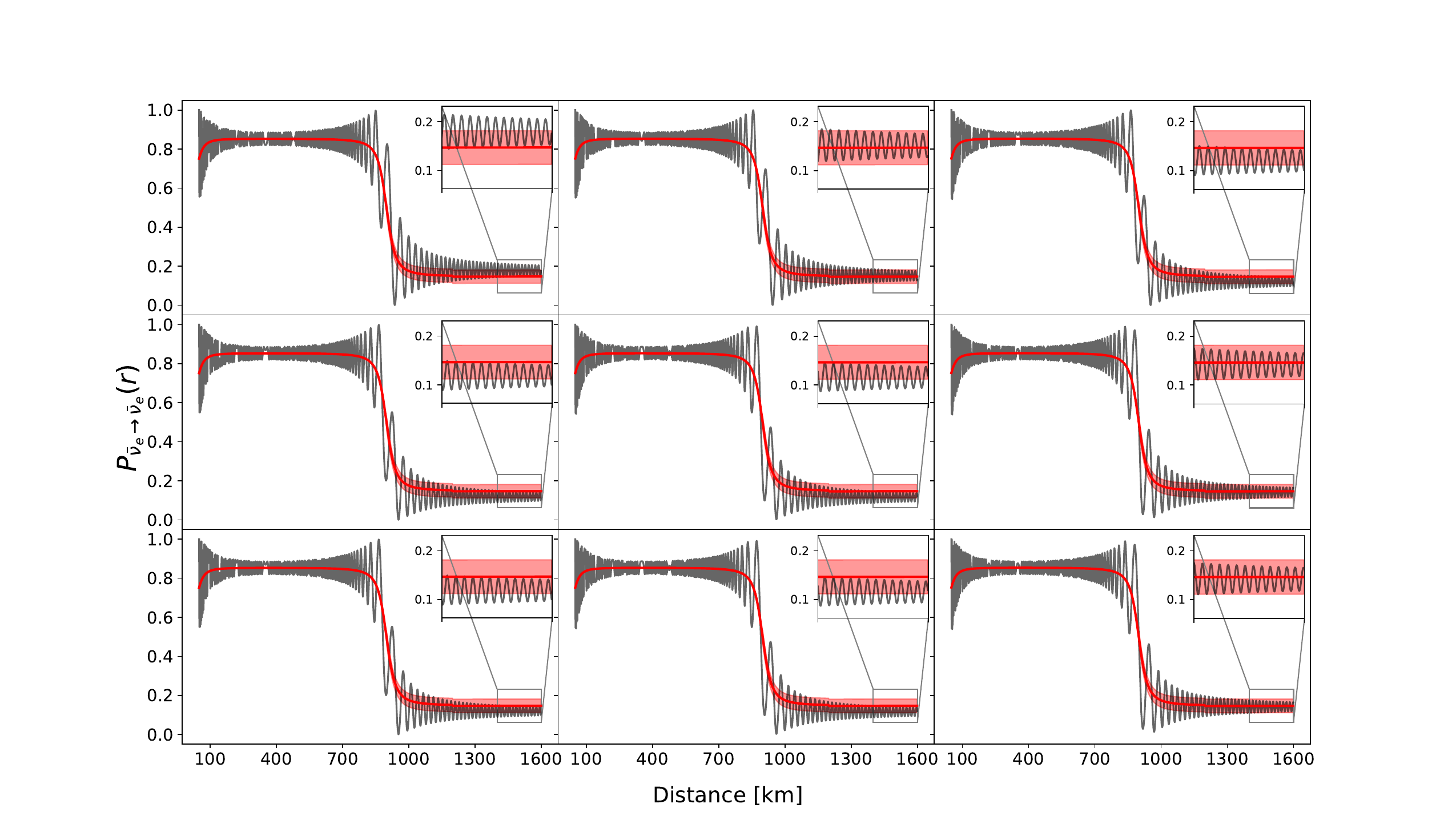}
\end{center}
\caption{
The survival probability of a $15$ MeV electron antineutrino at $5s$ post-bounce
time for exponentially fitted density distributions under slightly different
evolution conditions. The rows correspond to $R/\mbox{km}=49.95$, $50.00$,
$50.05$ and the columns correspond to $r_{\mbox{\footnotesize
mag}}/\mbox{km}=49.95$, $50.00$, $50.05$. The grey curves are numerical
solutions.  The thick red lines (same in every panel) show the theoretically
expected mean survival probability in the absence of phase effects (first two
lines of Eq.  (\ref{P ebar with SFP})). The red shaded regions (same in every
panel) show the theoretically expected range of mean survival probability with
the phase effects (the uncertainty from the third line of Eq.  (\ref{P ebar with
SFP}).) There is a partially adiabatic SFP resonance at $850$ km. Before that,
the survival probability oscillates around the thick red line indicating SFP
without phase effects. After the resonance, the survival probability oscillates
around a different mean value in each case which indicates phase effects.
However, in each panel the mean survival probability falls in the red shaded
region as expected.} 
\label{errors}
\end{figure*}

The adiabaticity of SFP resonance is quantified by \cite{Lim:1987tk,
1992PhR...211..167P, Nunokawa:1996gp} 
\begin{equation}
\label{gamma B}
\Gamma_B=\left(
\frac{(\mu B)^2} 
{\sqrt{2}G_F \frac{1}{m_n}\left\lvert\tfrac{dn(r)}{dr}\right\lvert(1-2Y_e)}
\right)_{\mbox{\footnotesize res}},
\end{equation}
where the subscript ``res'' indicates that the expression should be calculated
at the SFP resonance. The Landau-Zener approximation tells us that the probability
of the system to jump between the resonating energy eigenstates $\ket{r_3}$ and
$\ket{r_4}$ in NH is 
\begin{equation}
\label{Landau-Zener probability}
P_B=e^{-2\pi \Gamma_B}.
\end{equation}
Therefore the evolution is described by 
\begin{equation}
\label{Landau-Zener transfer}
\begin{pmatrix} 
\ket{r_3} \\ \ket{r_4} 
\end{pmatrix}
\to
\begin{pmatrix} 
\sqrt{1-P_B} & -e^{-i\alpha}\sqrt{P_B} \\ 
e^{-i\alpha}\sqrt{P_B} & \sqrt{1-P_B} 
\end{pmatrix}
\begin{pmatrix} 
\ket{r_3} \\ \ket{r_4} 
\end{pmatrix}
\end{equation}
through the resonance\footnote{Strictly speaking, Eqs. (\ref{gamma
B})-(\ref{Landau-Zener transfer}) assume that the magnetic field is constant
around the resonance region. This is not the case for our magnetic field
profile. But the comparison between our numerical and analytical results
indicate that this variation can be ignored here.}. Here $\alpha$ is called the
Stoke's phase \cite{PhysRevA.50.843, PhysRevA.55.R2495}. If $P_B\approx 0$ then
no jumping occurs from one eigenstate to the other which is the adiabatic limit.
In this limit the Stoke's phase becomes irrelevant. The Stoke's phase is also
irrelevant in the opposite (sudden) limit where $P_B\approx 1$ because it can be
included in the definition of the local eigenstates. But if $0<P_B<1$, then the
Stoke's phase should be included in the calculation. However, as we discuss
below, it becomes important only when the state entering the resonance is a
combination of the energy eigenstates $\ket{r_3}$ and $\ket{r_4}$. 

Neutrinos evolve adiabatically until the SFP resonance point, which we denote by
$r_B$. Before the resonance, the density operator is given by Eq. (\ref{later rho
in eigenbasis}). After the resonance, it is given by 
\begin{eqnarray}
\label{rho after SFP resonance}
\hat \rho (r)
&=&\!\!\rho_{11}(R)\ket{r_1}\bra{r_1} +\rho_{22}(R) \ket{r_2}\bra{r_2}\nonumber\\
&+&\!\! \left((1-P_B)\rho_{33}(R)+P_B\rho_{44}(R)\right)\ket{r_3}\bra{r_3}\nonumber\\
&+&\!\!\left(P_B\rho_{33}(R)+(1-P_B)\rho_{44}(R)\right)\ket{r_4}\bra{r_4}\\
&+&\!\!\left(e^{i\phi_B}\! \sqrt{P_B(1\!-\!P_B)} \rho_{34}(R) \! +
\!\mbox{cc}\right)\left(\ket{r_3}\bra{r_3}\!-\!\ket{r_4}\bra{r_4}\right)\nonumber\\
&+&\!\!(\dots)
e^{-i\int_{r_B}^r(E_3-E_4)dr} \ket{r_3}\bra{r_4} + \mbox{hc}\nonumber
\end{eqnarray}
in accordance with Eq. (\ref{Landau-Zener transfer}). Here, $\rho_{ij}(R)$ are
the matrix elements of the density operator in the energy eigenbasis at $R$
defined in Eq. (\ref{initial R in eigenbasis}). The last line of this result is
off-diagonal in the energy eigenbasis and fluctuates very fast around zero. Here
hc denotes the hermitian conjugate of this last term. The coefficients in the last line are not
shown explicitly because they will eventually die due to decoherence as
explained in the previous section. For this reason, we focus on the first four
lines which are diagonal. These terms change smoothly as the energy eigenstates
slowly vary with external conditions and determine the survival probability over
long distances. First three lines among them represent the ``classical''
outcome (i.e., no phase effects)
in the sense that they depend only on the Landau-Zener transition probabilities.
In contrast, the fourth line is an ``interference'' term (i.e., phase effects) which depends on the
relative phase between $\ket{r_3}$ and $\ket{r_4}$ acquired from the production
point $R$ to the SFP resonance point $r_B$, as well as the Stoke's phase $\alpha$. 
The sum of these two phases denoted by $\phi_B$ in the fourth line of
Eq. (\ref{rho after SFP resonance}), i.e.,  
\begin{equation}
\label{phi_B}
\phi_B=-\int_R^{r_B}(E_3(r)-E_4(r))dr +\alpha,
\end{equation}
is precisely what creates the SFP phase effect. 
Notice that this is a fixed phase, i.e., it does not lead to oscillations.
Instead, it 
contains the cumulative effects of the individual evolutions energy
eigenstate components from the production point
$R$ to the SFP resonance point $r_B$, giving rise to an interference between
them. 
Naturally, this occurs only when the terms that multiply
it in the fourth line are different from zero, i.e., when $\rho_{34}(R)\neq 0$
and $0<P_B<1$. If these conditions are satisfied then the interference term
affects the survival probabilities in $r\to\infty$ limit and creates the SFP
phase effect. The phase $\phi_B$ depends sensitively on the details of how the
neutrino undergoes from production point to the resonance point. But, as
explained at the beginning of Section III, neutrinos represented by
$\hat\rho(r)$ would be subject to slightly different evolution conditions. In
practice $\phi_B$ will be different for every neutrino and therefore it should be
treated as an \emph{uncertainty} by taking $-\pi\leq \phi_B< \pi$. 

The emergence of the SFP phase effect due to the phase given in Eq.
(\ref{phi_B}) and its sensitive dependence on external conditions are
demonstrated in Fig. \ref{errors}. This figure shows the survival probability of
a $15$ MeV electron antineutrino as a function of distance under slightly
different evolution conditions.  We use the exponentially fitted density
distribution at $5s$ post-bounce time, but change the external conditions by
slightly varying the radius of the neutrinosphere $R$ and the distance scale
$r_{\mbox{\footnotesize mag}}$ with which the magnetic field decreases. The rows
correspond to $R/\mbox{km}=49.95$, $50.00$, $50.05$ and the columns correspond
to $r_{\mbox{\footnotesize mag}}/\mbox{km}=49.95$, $50.00$, $50.05$. The
variation of $R$ mimics possible decoupling at slightly different radii from the
neutron star, or traveling slightly different distances due to being emitted at
different angles. The variation of $r_{\mbox{\footnotesize mag}}$ mimics
experiencing slightly different magnetic field profiles due to changing
conditions or traveling at different angles with it. 

The grey lines in Fig. \ref{errors} show the solutions obtained by numerically
solving the evolution equation given in Eq. (\ref{eqn:EoM}).  The thick red
lines show the theoretically expected average survival probabilities in the
absence of SFP phase effects. The fact that the grey lines are oscillating
perfectly around the thick red lines before $850$ km tells us that there are no
phase effects in this region. This is to be expected because, although the
neutrino fulfills the first condition for the emergence of the phase effects
mentioned in the Introduction (see below), the second condition is satisfied
only when it goes through the partially adiabatic SFP resonance which sits at
$850$ km. The red shaded regions show the range of the theoretically expected
average survival probability with the phase effects taken into account by
setting $-\pi\leq \phi_B< \pi$. The fact that the grey lines start oscillating
around averages which fall into the red shaded regions instead of oscillating
around the thick red lines after $850$ km shows that the SFP phase effect has
emerged. This can be better observed in the insets where the low density parts
are enlarged. 

It is useful to mention how we 
obtain the theoretically expected survival probabilities (thick red line and
the red shaded regions) in Fig. \ref{errors}. For this, we start from a pure ensemble as
in Eq. (\ref{box}) and use Eq. (\ref{local eigenkets SFP}) to obtain non-zero matrix
elements of the density matrix in energy eigenbasis: 
\begin{equation}
\label{rho(R) for initial ebar}
\begin{split}
&\rho_{33}(R)=\sin^2\!\theta_B(R), \quad \rho_{44}(R)=\cos^2\!\theta_B(R),\\
&\rho_{34}(R)=\rho_{43}(R)=\sin\theta_B(R)\cos\theta_B(R).
\end{split}
\end{equation}
The second line of the above equation is what tells us that the neutrino is
already in a superposition of energy eigenstates when it is born.  Eqs.
(\ref{general survival probability}), (\ref{later rho in eigenbasis}) and
(\ref{rho after SFP resonance}) then yield the following formula for survival
probability: 
\begin{align}
\label{P ebar with SFP}
P_{\bar\nu_e\to\bar\nu_e}\!(r)&\!=\!
\left[(1\!-\!P_B)\!\sin^2\!\theta_B(R)\!+\!P_B\!\cos^2\!\theta_B(R)\right]\abs{\bra{r_3}\ket{\bar\nu_e}}^2\nonumber\\
&+\!\left[P_B\!\sin^2\!\theta_B(R)\!+\!(1\!-\!P_B)\!\cos^2\!\theta_B(R)\right]\abs{\bra{r_4}\ket{\bar\nu_e}}^2\nonumber\\
&\!\!\!\!\!\!\pm\!\sqrt{P_B(1\!-\!P_B)}\sin
2\theta_B(R)(\abs{\bra{r_3}\ket{\bar\nu_e}}^2\!-\!\abs{\bra{r_4}\ket{\bar\nu_e}}^2)\nonumber\\
&\!\!\!\!\!\!+\mbox{terms oscillating around zero}.
\end{align}
This formula is valid with $P_B=0$ before the SFP resonance at $850$ km, and
with $P_B$ given by Eq. (\ref{Landau-Zener probability}) after it.  The last
line contains fast oscillations around zero which we do not explicitly write
down. If we disregard these fast oscillations, what remains is the mean survival
probability described by the first three lines.  The first two lines come from
the classical probability (i.e., no phase effects) part of Eq.  (\ref{rho after
SFP resonance}) and give us the thick red line in Fig. \ref{errors}. The third
line comes from interference (i.e., phase effect) term of Eq.  (\ref{rho after
SFP resonance}). Varying the phase within the full range as explained above
brings the $\pm$ sign and gives us the red shaded uncertainty region around the
thick red line. Note that the small variations of $R$ and
$r_{\mbox{\footnotesize mag}}$ mentioned above are almost irrelevant for the
values of $P_B$ and $\theta_B(R)$. For this reason, the thick red line and the
red shaded regions are identical in all the panels of Fig. \ref{errors}

\subsection{At \texorpdfstring{$r\to\infty$ limit}{r to infinity limit}}

The decoherence over long distances can be implemented by discarding the
off-diagonal terms of the density operator given in Eq. (\ref{rho after SFP
resonance}), which removes all oscillations. Using Eq. (\ref{eigenstates in
vacuum}) for NH we find the following analytical expression for the density
operator in $r\to\infty$ limit:
\begin{eqnarray}
\label{rho at infinity after SFP}
\hat\rho(\infty)\!\!
&=&\!\!\rho_{11}(R)\ket{\bar\nu_2}\bra{\bar\nu_2} +\rho_{22}(R) \ket{\nu_2}\bra{\nu_2}\\
&+&\!\!\left((1-P_B)\rho_{33}(R)+P_B\rho_{44}(R)\right)\ket{\nu_1}\bra{\nu_1}\nonumber\\
&+&\!\!\left(P_B\rho_{33}(R)+(1-P_B)\rho_{44}(R)\right)\ket{\bar\nu_1}\bra{\bar\nu_1}\nonumber\\
&\pm&\!\! 
2\sqrt{P_B(1-P_B)}\abs{\rho_{34}(R)}(\ket{\nu_1}\bra{\nu_1}-\ket{\bar\nu_1}\bra{\bar\nu_1}).\nonumber
\end{eqnarray}
In particular, we find the analytical expression for the limiting survival
probability of an initial $\bar\nu_e$ from Eqs. (\ref{rho(R) for initial ebar})
and (\ref{rho at infinity after SFP}) as 
\begin{align}
\label{P ebar -> infinity with SFP}
P_{\bar\nu_e\to\bar\nu_e}(\infty)
&=\left(P_B\sin^2\!\theta_B(R)+(1-P_B)\cos^2\!\theta_B(R)\right)\nonumber\\
&\pm\sqrt{P_B(1-P_B)}\sin 2\theta_B(R).
\end{align}
Here we also use Eq. (\ref{eigenstates in vacuum}) and the fact that 
$\bra{\nu_1}\ket{\bar\nu_e}=0$ and $\bra{\bar\nu_1}\ket{\bar\nu_e}=1$ since the
mixing angle is taken to be zero. 

In our numerical simulations the oscillations do not die over long
distances because our equation of motion given in Eq. (\ref{eqn:EoM}) does not
take decoherence into account. We approach this situation as follows: First, we
run the simulation until the density is low enough to be considered vacuum. This
happens somewhere between a few to several thousand kilometers depending on
the post-bounce time and the neutrino energy. Once we observe that the survival
probability starts to oscillate steadily around a fixed average value, we stop
the simulation and remove the off diagonal elements of the resulting density
operator in mass basis. Doing this brings the survival probability to the
average value around which the numerical result steadily oscillates in vacuum. 

\begin{figure}
\begin{center}
\includegraphics[width=0.92\columnwidth]{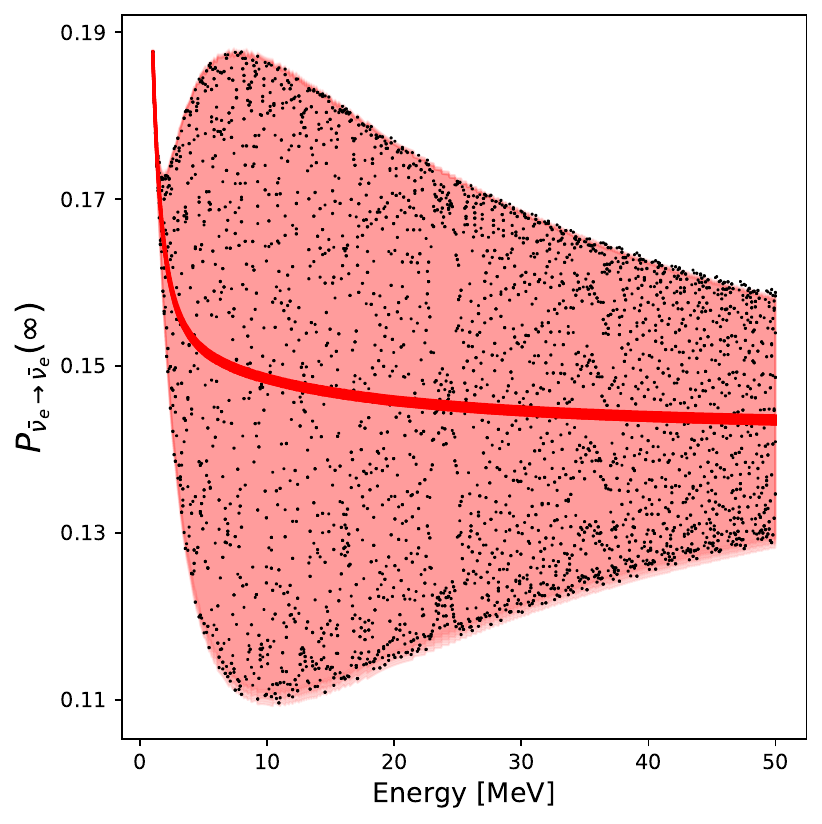}
\end{center}
\caption{The survival probability of a $\bar\nu_e$ for the exponentially fitted
density distribution at $t=5$ s in $r\to\infty$ limit as a function of energy.
The thick red line is the classical probability result obtained from the first
line of Eq. (\ref{P ebar -> infinity with SFP}). The red shaded region is the
uncertainty obtained from the second line of the same equation. For each
neutrino energy, $9$ numerical simulations are carried out with the same $R$ and
$r_{\mbox{\footnotesize mag}}$ parameters as in Fig. \ref{errors}. The black
dots represent the limiting survival probabilities obtained from these
simulations as described in the text.} 
\label{errors_energy}
\end{figure}
We show the $\bar\nu_e$ survival probability at $r\to\infty$ limit as a function
of energy for the exponentially fitted density distribution at $t=5$ s in Fig.
\ref{errors_energy}. The solid red line is the classical probability result
obtained from the first line of Eq. (\ref{P ebar -> infinity with SFP}) and the
red shaded region is the uncertainty from the second line of the same equation.
Each black dot in this figure represents a numerical run. To obtain them, we
divide the energy range into $250$ bins, and run $9$ simulations for each bin
with the same $R$ and $r_{\mbox{\footnotesize mag}}$ parameters as those in Fig.
\ref{errors}. But unlike in Fig. \ref{errors} where we stop at $1600$ km, in
Fig. \ref{errors_energy} we continue the simulations until the vacuum is reached
and obtain the numerical value of the limiting survival probability as described
above. The result of each run is shown with a black dot in Fig.
\ref{errors_energy}. Note that, if we used the same set of external conditions
for each energy bin, the survival probability would display fast oscillations
with energy. But, since even those neutrinos with the same energy are likely to
experience slightly different external conditions as discussed above, we choose
this approach and present our results as scattering plots as in Ref. \cite{Fogli:2003dw}.
A total of $2250$ numerical simulations were
carried out to generate this figure and using exponential fits for the density
profiles substantially reduces the total running time. Fig. \ref{errors_energy} tells
us that the survival probability of an initial $\bar\nu_e$ can change by as much
as $\%30$ depending on the energy under the adopted conditions. But in each
numerical run the limiting survival probability always falls into the
uncertainty range predicted by Eq. (\ref{P ebar -> infinity with SFP}).

\section{Non-zero mixing angle}

When the mixing angle is not zero, neutrinos go through both SFP and MSW
resonances. For $Y_e>1/3$, which is most often the case, MSW resonance takes
place after the SFP resonance. The adiabaticity of the MSW resonance is
quantified by the parameter
\begin{equation}
\label{gamma M}
\Gamma_M=\left(
\frac{(\tfrac{\delta m^2}{2E}\sin 2\theta)^2} 
{\sqrt{2}G_F \frac{1}{m_n}\left\lvert\tfrac{dn(r)}{dr}\right\lvert Y_e}
\right)_{\mbox{\footnotesize res}},
\end{equation}
where the subscript ``res'' indicates that the expression should be calculated
at the resonance. The Landau-Zener approximation tells us that the jumping probability
between the resonating energy eigenstates $\ket{r_2}$ and $\ket{r_3}$ in NH is
\begin{equation}
\label{Landau-Zener probability MSW}
P_M=e^{-2\pi \Gamma_M}.
\end{equation}
Therefore the evolution through the MSW resonance is described by 
\begin{equation}
\label{Landau-Zener transfer MSW}
\begin{pmatrix} 
\ket{r_2} \\ \ket{r_3} 
\end{pmatrix}
\to
\begin{pmatrix} 
\sqrt{1-P_M} & -e^{-i\beta}\sqrt{P_M} \\ 
e^{-i\beta}\sqrt{P_M} & \sqrt{1-P_M} 
\end{pmatrix}
\begin{pmatrix} 
\ket{r_2} \\ \ket{r_3} 
\end{pmatrix},
\end{equation}
where $\beta$ is the Stoke's phase. For a completely adiabatic ($P_M=0$) or
completely nonadiabatic ($P_M=1$) resonance the Stoke's phase is unimportant.
But in partially adiabatic case with  $0<P_M<1$, it should be taken into
account.

In between the SFP and MSW resonances the density operator is given by Eq.
(\ref{rho after SFP resonance}). After the MSW resonance Eq. (\ref{Landau-Zener
transfer MSW}) must also be applied to it which results in a complicated generic
form. Here we do not reproduce the full result because it is not relevant for
our purposes. Instead, we give its $r\to\infty$ limit which we obtain by
removing its off-diagonal components and using Eq. (\ref{eigenstates in vacuum}). 
The result is 
\begin{align}
	\label{rho at infinity theta}
	\nonumber\hat \rho(\infty)=&\rho_{11}(R)\ket{\bar\nu_2}\bra{\bar\nu_2}\\
	\nonumber&+[(1\!-\!P_M)\rho_{22}(R)\!+\!P_M((1\!-\!P_B)\rho_{33}(R)\!+\!P_B\rho_{44}(R))]\\
	\nonumber& \qquad\qquad \times\ket{\nu_2}\bra{\nu_2}\\ 
	\nonumber&+[P_M \rho_{22}(R)\!+\!(1\!-\!P_M)((1\!-\!P_B)\rho_{33}(R)\!+\!P_B\rho_{44}(R))]\\
	\nonumber& \qquad\qquad \times\ket{\nu_1}\bra{\nu_1}\\
	\nonumber&+[P_B \rho_{33}(R)\!+\!(1\!-\!P_B) \rho_{44}(R) ]\ket{\bar\nu_1}\bra{\bar\nu_1}\\
	\nonumber&\pm 2\sqrt{(1\!-\!P_B) P_B}\,\abs{\rho_{34}(R)}\\
	\nonumber& \qquad\qquad \times[(1\!-\!P_M)\ket{\nu_1}\bra{\nu_1}\!+\!P_M\ket{\nu_2}\bra{\nu_2}\!-\!\ket{\bar\nu_1}\bra{\bar\nu_1}]\\
	\nonumber&\pm 2\sqrt{(1\!-\!P_M)P_M}[\sqrt{P_B}\,\abs{\rho_{24}(R)}\!+\!\sqrt{1\!-\!P_B}\,\abs{\rho_{23}(R)}]\\
	& \qquad \qquad \times (\dyad{\nu_2}{\nu_2}\!-\!\dyad{\nu_1}{\nu_1})
\end{align}
The first four lines of this equation involve only the probabilities and
represent the classical result. The last two lines give the uncertainty due to
the phases. In addition to the one given in Eq. (\ref{phi_B}), other phases also
enter the result: These are the relative phases acquired by $\ket{r_3}$ and by $\ket{r_4}$
with respect to $\ket{r_2}$ from production to the SFP resonance, the relative
phase between $\ket{r_2}$ and $\ket{r_3}$ from the SFP resonance to the MSW
resonance, and the Stoke's phase $\beta$ associated with the MSW resonance.
These phases show up in combinations so that there are only two uncertainty
terms in the final result. Note that Eq. (\ref{rho at infinity theta}) reduces
to Eq. (\ref{rho at infinity after SFP}) when the MSW resonance is fully
adiabatic, i.e., when $P_M=0$. In particular, for an initial electron
antineutrino, the limiting survival probability can be calculated by using Eq.
(\ref{rho(R) for initial ebar}) and the relations $\bra{\nu_1}\ket{\bar\nu_e}=0$
and $\bra{\bar\nu_1}\ket{\bar\nu_e}=\cos\theta$. The result is 
\begin{align}
\label{rho ebar -> infinity theta}
P_{\bar\nu_e\to\bar\nu_e}(\infty)
&\!=\!\left(P_B\sin^2\!\theta_B(R)\!+\!(1\!-\!P_B)\cos^2\!\theta_B(R)\right)\cos^2\theta \nonumber\\
&\pm\sqrt{P_B(1\!-\!P_B)}\sin 2\theta_B(R)\cos^2\theta, 
\end{align}
which reduces to Eq. (\ref{P ebar -> infinity with SFP}) when $\theta=0$. This
formula does not contain $P_M$. This is expected because we focus on NH in which
case antineutrinos do not experience the MSW resonance. However, from Fig.
\ref{eigenvalues} we see that the component of the initial $\bar\nu_e$ which
turns into $\nu_x$ in SFP resonance later experiences MSW resonance, and here it can
partially or fully turn into $\nu_e$. Therefore we expect the
$\bar\nu_e\to\nu_e$ transition probability to involve both $P_B$ and $P_M$.
Substituting Eq. (\ref{rho(R) for initial ebar}) into Eq. (\ref{rho at infinity
theta}) and taking the matrix element $\mel{\nu_e}{\hat\rho(\infty)}{\nu_e}$
leads to the result 
\begin{align}
\label{rho ebar -> e infinity theta}
&\begin{multlined}[t][0.95\columnwidth]
P_{\bar\nu_e\to\nu_e}(\infty)
=\left(P_B\sin^2\!\theta_B(R)+(1-P_B)\cos^2\!\theta_B(R)\right)\\
\times\left((1-P_M)\cos^2\theta+P_M\sin^2\theta\right) 
\end{multlined}
\nonumber\\
&
\quad\quad
\quad\quad
\quad
\pm\sqrt{P_B(1\!-\!P_B)}\sin 2\theta_B(R)\cos^2\theta 
\end{align}
\begin{figure}[hbt!]
\centering
\includegraphics[height=0.65\textheight]{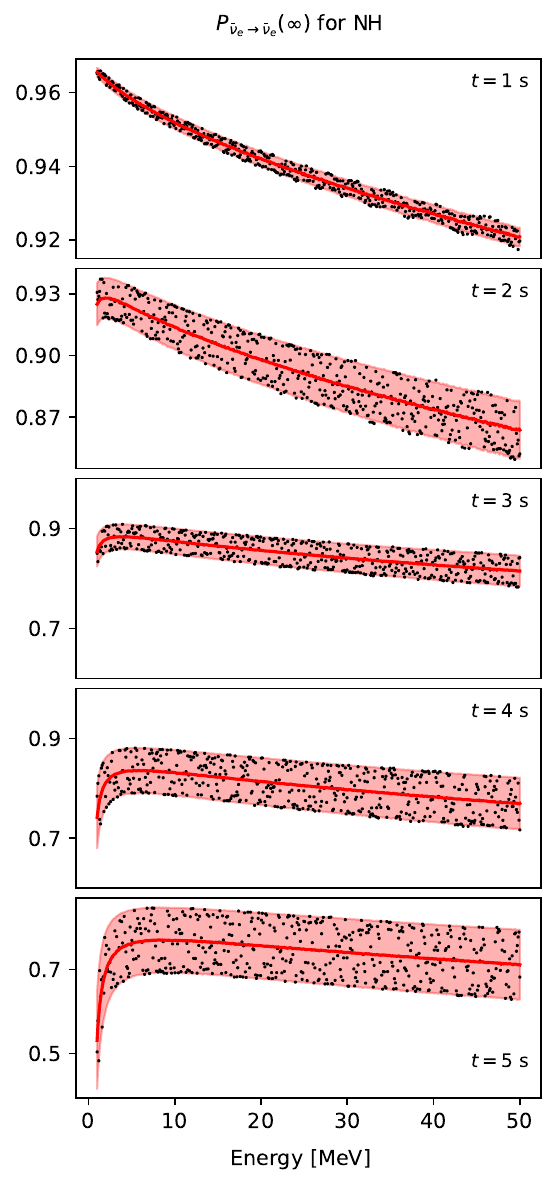}
\caption{The electron antineutrino survival probability at $r\to\infty$ limit as a
function of energy for $\mu=1\times 10^{-16} \mu_B$ and for exponentially
fitted density distributions. The panels show the
post-bounce times $t=1,2,3,4,5$ s from top to bottom. The solid red lines and
the red shaded regions respectively show the average survival probability and
the associated uncertainty calculated from Eq. (\ref{rho ebar -> infinity
theta}). The black dots are the limiting survival probabilities obtained from
numerical solutions under slightly different and randomly chosen conditions as
described in the text.}
\label{random initial for 1e16}
\end{figure}

\begin{figure}[hbt!]
\centering
\includegraphics[height=0.65\textheight]{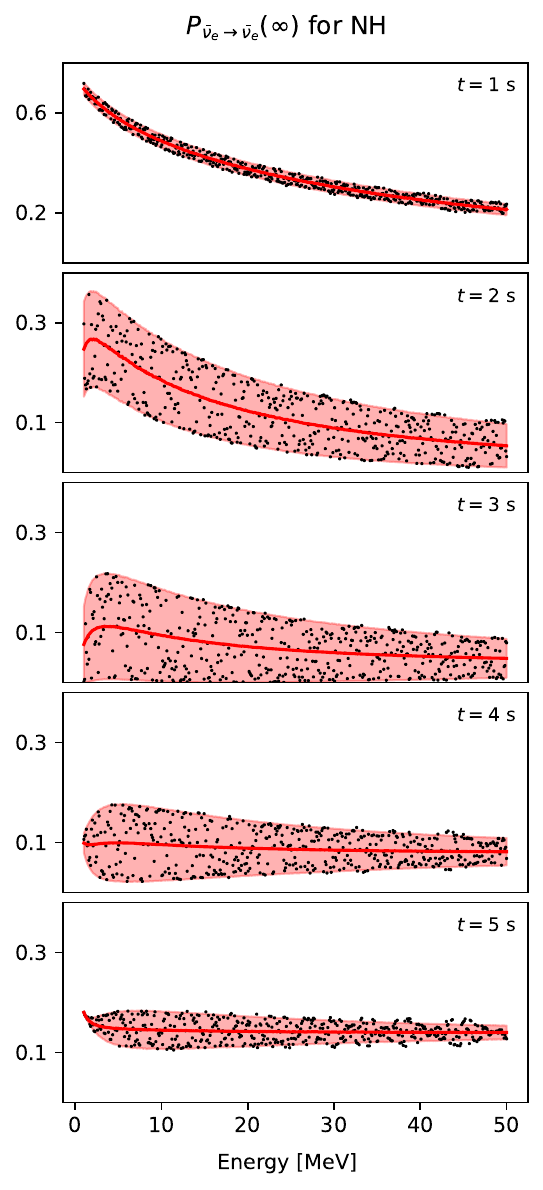}
\caption{Same as Fig. \ref{random initial for 1e16}, but for $\mu=5\times
10^{-16} \mu_B$.} 
\label{random initial for 5e16}
\end{figure}
\begin{figure}[hbt!]
\centering
\includegraphics[height=0.65\textheight]{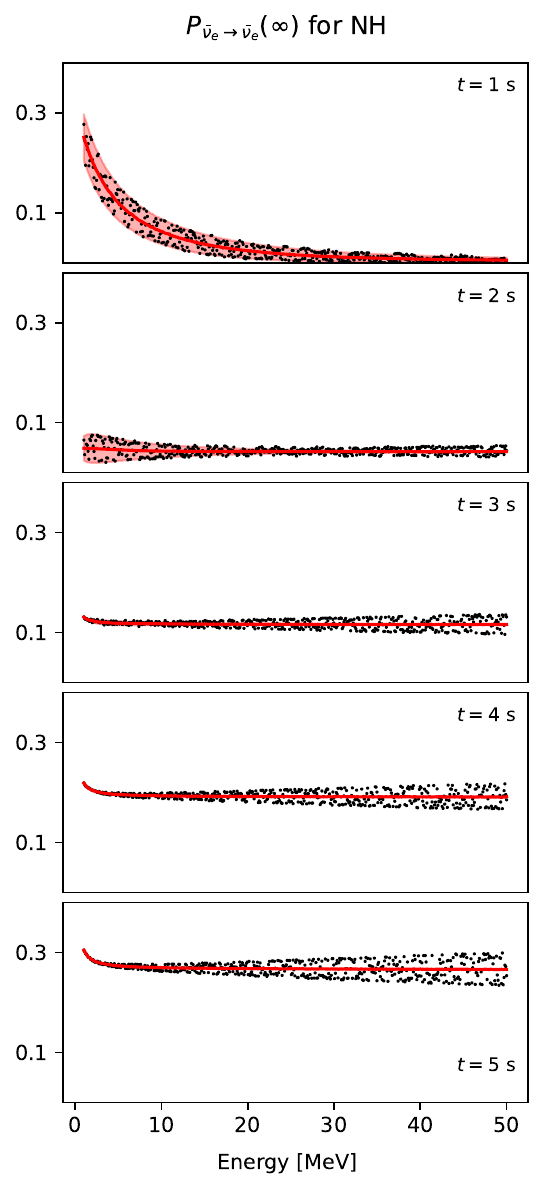}
\caption{Same as Fig. \ref{random initial for 1e16}, but for $\mu=10\times
10^{-16} \mu_B$.} 
\label{random initial for 1e15}
\end{figure}
Fig. \ref{random initial for 1e16} shows the survival probability of an electron
antineutrino at $r\to\infty$ limit as a function of neutrino energy for
$\mu=1\times 10^{-16}\mu_B$ and for the exponentially fitted density
distributions at different post-bounce times. The solid red lines
show the average survival probability and the red shaded areas show the uncertainty
region, both calculated from Eq. (\ref{rho ebar -> infinity theta}). The black
dots represent the results obtained by numerically solving Eq. (\ref{eqn:EoM})
under very similar but slightly different conditions as described in the context
of Fig. \ref{errors_energy}. But this time, for each neutrino energy we solve
the evolution equation three times. In each run we choose both $R$ and
$r_{\mbox{\tiny mag}}$ randomly and independently in the interval $(49.95 \mbox{
km}, 50.05 \mbox{ km})$. As can be seen, all numerical results are within the
uncertainty bounds predicted by Eq. (\ref{rho ebar -> infinity theta}). 

Fig. \ref{random initial for 5e16} shows the same results for $\mu=5\times
10^{-16}\mu_B$. Here we see a few differences from Fig. \ref{random initial for
1e16}. First, the survival probability of $\bar\nu_e$ drops to significantly
lower values. This is because for larger $\mu$ the SFP resonance is more
adiabatic according to Eq. (\ref{gamma B}), which makes the $\bar\nu_e \to \nu_x$
transition more efficient. Second, some numerical results are slightly out of
the analytical uncertainty region at late times. This is because our analytical
treatment is based on the decoupling approximation in which SFP and MSW
resonances can be treated as two separate two-level problems within the
Landau-Zener formalism. Due to the widening of SFP resonance examined in Section
III, this approach eventually fails at late times, especially for larger $\mu B$
values.

Fig. \ref{random initial for 1e15} shows our results for an even larger magnetic
moment of $\mu=10\times 10^{-16} \mu_B$. In this figure two things can be
observed: First, the uncertainty predicted by our analytical approach becomes
zero at late post-bounce times. This is because the SFP moves inside and happens
under a stronger magnetic field. Eventually, the $\mu B$ value at the SFP resonance becomes large enough
to render it completely adiabatic at late times. MSW resonance is
also adiabatic for the model that we work with. In this case, the predicted
uncertainty vanishes. In other words since both resonances are adiabatic
analytical approach suggests that there is no phase effect. However, (and this
is the second observation) we see that the numerical results still indicate the
presence of a phase effect. This is due to the failure of the decoupling scheme.
As discussed in Section III, the SFP resonance becomes wider at late post-bounce
times. This widening effect is even more pronounced for large $\mu$. We
conclude that if SFP and MSW resonances overlap, the phase effect may still be
there even when individual resonances seem to be adiabatic from a
straightforward application of the Landau-Zener approach. 

\section{Observability of SFP Phase Effect}

Phase effects cause the neutrino survival and transition probabilities to
randomly fluctuate around an average value. From an observational point of view, each dot in Figs.
\ref{random initial for 1e16} - \ref{random initial for 1e15} may be thought as
a possible neutrino interacting at the detector. If many neutrinos are captured
in a given energy bin in a given time interval, their combined effect would
average out this fluctuation. 
In other words, although neutrinos emitted in a small time and energy interval arrive at the
detector in different final states, the net effect would be the
same as if the detector is responding to a smooth neutrino signal. In
this case, it is appropriate to ignore the phase effects from the beginning as
is often the practice in the literature. But, if only a few
neutrinos can be captured in each energy bin, then averaging cannot completely
remove the randomization. In this case, the detector would in principle be
responding to an irregular neutrino signal. However, this irregularity can still get
lost if the detector cannot resolve it. In what follows, we
first demonstrate SFP phase effect in a realistic scenario, and then talk about
its observability. 

We start with the mixed ensemble of neutrinos as described in Eq.
(\ref{initial R in flavor basis}). We take the initial energy distributions to
be of Fermi-Dirac type\footnote{If the neutrino magnetic moment is large, then
the neutrino spectra emerging from the proto-neutron star may be different from
Fermi-Dirac type due to neutrino electromagnetic interactions inside the core.
See, for example, Refs. \cite{Dar:1987yv, Kuznetsov:2009we, Alok:2022ovy}. We
do not consider this effect here.} and evolve it using the realistic density
profiles, i.e., the solid lines shown in Fig.  \ref{fig:baryonProfileShock}. As
discussed just below Eq. (\ref{fits}), the realistic
density profiles decrease more slowly than the analytical fits and consequently the
resonances take place in outer regions where the magnetic field is weaker. As a
result, magnetic moments of the order of $10^{-16} \mu_B$ used in illustrative
discussion create smaller amounts of smearing with realistic density profiles.
However, a stronger magnetic moment can offset this effect and create a larger
smearing. 

\begin{figure}[hbt!]
\centering
\includegraphics[width=0.80\columnwidth]{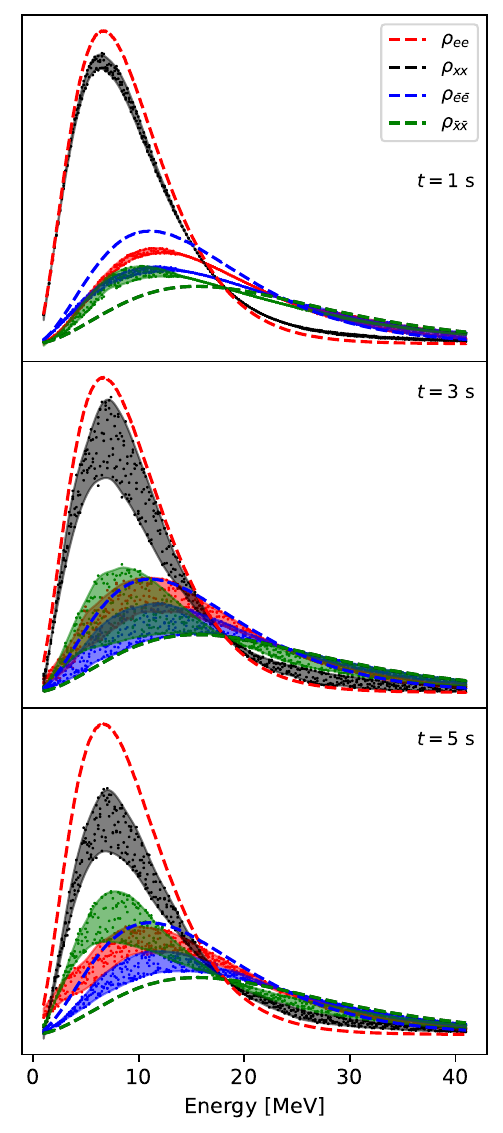}
\caption{Smearing of the supernova neutrino flux at Earth for $\mu=5\times
10^{-15} \mu_B$ in arbitrary units. The panels correspond to $t=1,3,5$ s from
top to bottom. The dashed lines show initial neutrino fluxes with Fermi-Dirac
type distributions. The dots in the same color coding show the numerical results
obtained by introducing small random variations for each energy bin as described in the text.
We use
realistic density profiles shown with the solid lines in Fig.
\ref{fig:baryonProfileShock}. The shaded areas do not represent the
analytically expected uncertainty as in previous figures, but serve to indicate
the distribution of numerical results.}
\label{fermi dirac}
\end{figure}
Fig. \ref{fermi dirac} shows our results for $\mu=5\times10^{-15} \mu_B$ with
realistic density profiles for a supernova at $10$ kpc distance from Earth with
a total energy of $3\times 10^{53}$ ergs. The energy is shared 
equally among all six flavors (see, e.g., Ref.
\cite{Fogli:2003dw}) while 
each flavor has a different
thermalization with $kT_{\nu_e}=3.0$ MeV,
$kT_{\bar\nu_e}=5.0$ MeV, and $kT_{\nu_x}=kT_{\bar\nu_x}=7.0$ MeV where
$T_{\nu_\alpha}$ denotes the temperature for the $\nu_\alpha$ flavor and $k$
denotes the Boltzmann constant.
The panels in this figure show post-bounce
times $t=1,3,5$ s from top to bottom. The dashed lines show the 
energy distribution of neutrino fluxes on the 
proto-neutron star surface with
red, black, blue, and green colors respectively
corresponding to $\nu_e,\nu_x,\bar\nu_e$, and $\bar\nu_x$ flavors. 
We use arbitrary units so as to be able to compare these fluxes with the
ones reaching Earth.
The dots represent the results of our numerical simulations. We obtain them by
solving the evolution described by Eq. (\ref{eqn:EoM}) for the realistic density
profiles up until the vacuum and then by applying decoherence in $r\to\infty$
limit. As before, we do this three times for each energy bin by choosing $R$ and
$r_{\mbox{\tiny mag}}$ randomly and independently in the interval $(49.95 \mbox{
km}, 50.05 \mbox{ km})$. If a single set of parameters were used instead
of randomization, the resulting energy spectra would rapidly oscillate around
an average distribution. However, as discussed before, it is more realistic to
assume that neutrinos would experience slightly different external conditions
due to their direction of travel, for instance. Unlike in previous figures, the
shaded regions in Fig. \ref{fermi dirac} do not represent analytically
calculated uncertainty ranges because we are using realistic density profiles
rather than their analytical fits. Instead, these regions are drawn by hand to
guide the eye about the spreading of the points.  Color coding for the dots and
the shaded regions are the same as those for the initial distributions.  Note
that, we calculate up to $100$ MeV neutrino energy but we show the spectra only
up to $40$ MeV in Fig. \ref{fermi dirac} for clarity. For larger energies, the
colored regions overlap with each other. However, high energy neutrinos are
important for the detector response. For this reason energies up to $100$ MeV
will be included in event rate calculations below.

Fig. \ref{fermi dirac} tells us that the phase effect makes little impact at the
early time of $t=1$ s. This is expected because the matter density at the
neutrinosphere is relatively high and flavor eigenstates are close to energy
eigenstates. However, at later times decreasing density near the neutrinosphere
drives the initial energy eigenstates away from flavor eigenstates and cause
neutrinos to experience a more pronounced SFP phase effect.  
Colored regions are particularly wide around $5-15$ MeV. This is due to the fact
that the original distributions differ most from each other in this energy
region and therefore any uncertainties in the survival and transition
probabilities make the strongest impact here. 

What is the significance of Fig. \ref{fermi dirac} from an observational point
of view?  The answer is that the colored regions are the zones in which the
energy distributions of the neutrinos \emph{captured by the detector} can
possibly lie. As discussed above, if the detector captures many neutrinos in
each energy bin, then the randomization will be removed through averaging and
the energy distributions will be smooth lines centered inside each colored
region. The result would be as if there is no phase effect. But, if a relatively
low number of neutrinos are captured, then the distributions would be randomized
inside the colored regions. The weaker the signal, the more randomized a
distribution the detector will be responding to. A supernova may fall into the
second category at late post-bounce times. A typical next generation neutrino
detector such as DUNE or Hyper
Kamiokande is expected to capture a few thousand neutrinos from a galactic core
collapse supernova. But most of them will be detected in the first few seconds
after the core bounce. The neutrino luminosity of a supernova decreases
approximately as $\exp(-t/3\;\mbox{s})$. For this reason, the detection rate
around $t=5$ s would drop to a few hundred per second. At the same time our
results indicate that the SFP phase effect becomes prominent in amplitude a few
seconds after the core bounce.  Therefore, if the SFP phase effect is detectable
at all, it is reasonable to assume that it will appear at late post-bounce
times. In what follows, we investigate the prospect of observing the SFP phase
effect with DUNE far detector for the example provided in Fig. \ref{fermi dirac}
at $t=5$ s. 

DUNE is a liquid scintillator detector currently under construction and it can
detect neutrinos in one of the following channels \cite{DUNE:2020zfm}:
\begin{equation}
	\label{channels}
\begin{split}
	\nu_e + \ce{^40Ar} &\longrightarrow \ce{^40K}^* + e^- \\
	\bar\nu_e + \ce{^40Ar} &\longrightarrow \ce{^40Cl}^* + e^+\\ 
\end{split}
\end{equation}
For these reactions, the observables are the outgoing leptons
together with the decay line of the final excited nuclei. 
All types of neutrinos can also scatter from
both Ar nuclei and electrons, but we do not consider these channels here. Scattering from Ar nuclei 
proceeds through neutral current interactions alone
and does not distinguish between different neutrino degrees of freedom. As a result, it is insensitive to 
flavor transformations. Scattering from electrons
in principle distinguishes $\nu_e$ and $\bar\nu_e$ which have
larger cross sections than other flavors, but we find that this is not enough to resolve the phase effects.
\begin{figure*}
\centering
\includegraphics[width=\textwidth]{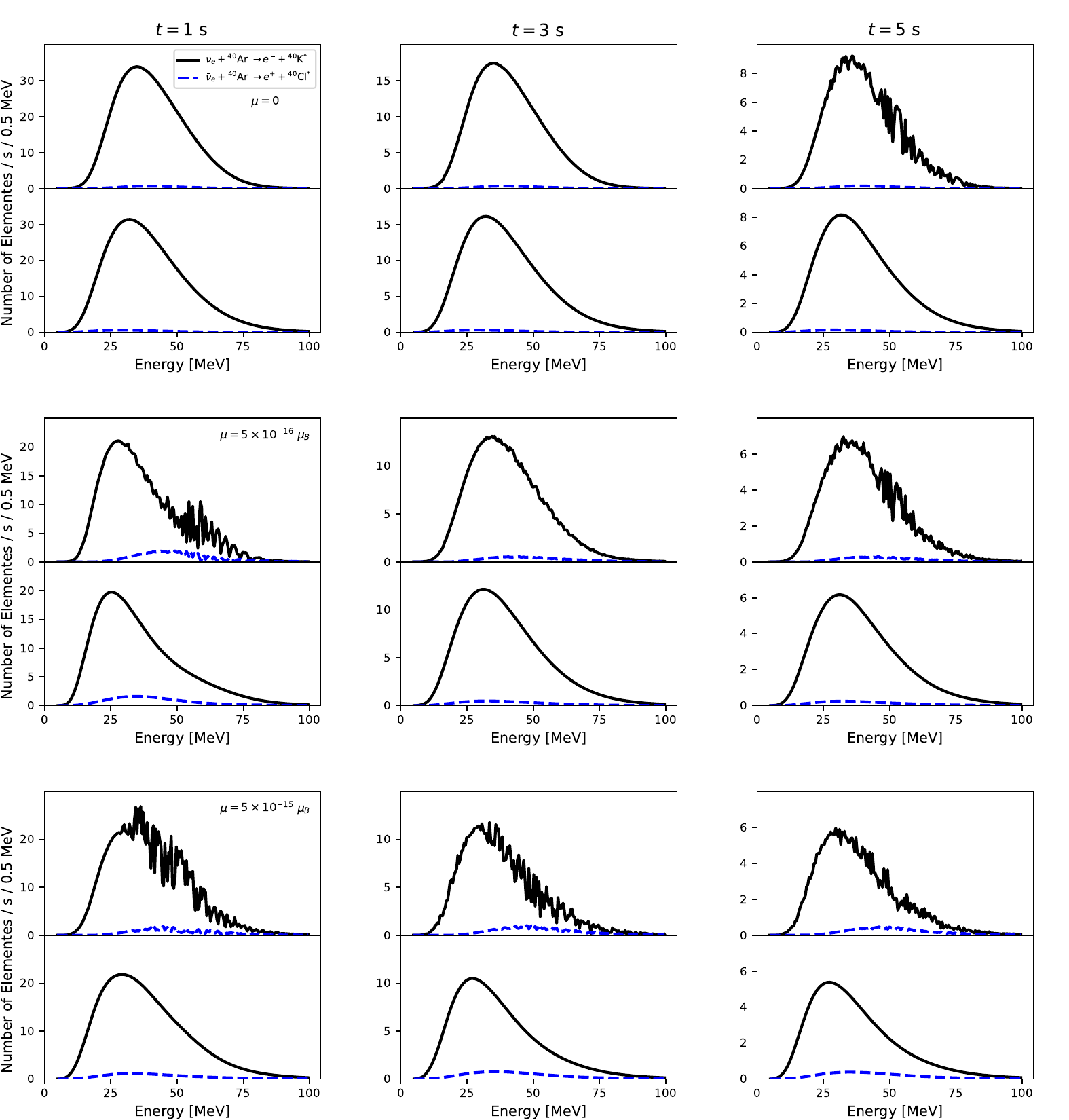}
\caption{Reaction rates on Ar nuclei in DUNE for $\mu=0$, $\mu=5\times10^{-16}
\mu_B$, and $\mu=5\times10^{-15} \mu_B$ (from top to bottom) at $t=1,3,5$ s
post-bounce time (from left to right) for the same model as in Fig. \ref{fermi
dirac}. In each plot, upper panels show the interaction rates as functions of
neutrino energy, assuming that it can be perfectly reconstructed (i.e., perfect
energy resolution). Lower panels show the same event rates as functions of
detected lepton energy after incorporating detector energy resolution as
described in the text.}
\label{event rates}
\end{figure*}

In Fig. \ref{event rates}, we show the rates of the Ar reaction channels
obtained with the SNOwGLoBES \cite{Huber:2004ka, Huber:2007ji,snowglobes} event
rate calculator for three different values of neutrino magnetic moment ($\mu=0$,
$\mu=5\times10^{-16} \mu_B$, and $\mu=5\times10^{-15} \mu_B$ from top to bottom)
at three different post-bounce times ($t=1,3,5$ s from left to right).  To
obtain the event rates for $\mu=5\times10^{-15} \mu_B$, we randomly choose one
dot for each flavor per $0.5$ MeV from the panels of Fig. \ref{fermi dirac}
(this time we use our full energy range up to $100$ MeV) and feed the resulting
irregular neutrino spectra into the event rate calculator. For other $\mu$
values, we calculate similarly.  The upper panels show the resulting interaction
rates as functions of the neutrino energy assuming that it can be perfectly
reconstructed, i.e., before the finite detector energy resolution is
incorporated.  As expected, the resulting event rates randomly fluctuate. The
fluctuations that we observe for $\mu=0$ are due to the ordinary phase effect
associated with the MSW resonances. Ordinary phase effect can appear only at
late times because it requires the presence of two MSW resonances which, as can
be seen in Fig. \ref{fig:baryonProfileShock}, happens only when the reverse
shock creates a dip in density profile. This tells us that, the fluctuations
that appear at earlier times for $\mu\neq 0$ are due to SFP phase effects. At
later times, ordinary and SFP phase effects are intertwined. According to Fig.
\ref{fermi dirac} for $\mu=5\times10^{-15} \mu_B$, SFP phase effect smearing is
smaller at $t=1$s in comparison to $t=3$s, but in the corresponding (i.e., the
last) row of Fig. \ref{event rates}, we observe larger event rate fluctuations
at $t=1$s than at $t=3$s. This is due to the fact that neutrino flux is stronger
at earlier times which gives rise to more events at the detector.  Also note
that in Fig. \ref{event rates} event rate fluctuations are observed only for
energies higher than about $25$ MeV despite the fact that Fig.  \ref{fermi
dirac} shows SFP phase effect smearing in lower energies as well.  This is due
to the decrease of neutrino cross sections with energy. The lower panels of Fig.
\ref{event rates} show the same event rates as a functions of detected lepton
energy after incorporating the finite detector energy resolution\footnote{This
is called detector energy smearing in the literature, but we avoid using this
term since we use the word smearing to describe the widening of neutrino energy
spectra due to the phase effects.}. This process involves folding the event
rates with the detector energy resolution function, which  is a Gaussian with
energy dependent standard deviation given by $\sigma(E) = E\;(
(0.11/\sqrt{E/\mbox{MeV}})^2+(0.02)^2)^{1/2}$ for an argon detector
\cite{snowglobes,ICARUS:2003zvt}.  We assume $100\%$ detection efficiency. The
fact that no fluctuations are observed in the lower panels tells us that, for
the supernova model that we employ and the parameters that we use, the SFP phase
effect is washed out in the final lepton spectra.  Although SFP phase effects
are washed out, SFP itself still makes an
impact on the observed event rates. This can be seen by comparing the first row
of Fig. \ref{event rates} ($\mu= 0$) with its second and third rows ($\mu\neq
0$).  For all post-bounce seconds, a decrease of electron events and a
slight enhancement of positron events can be observed for non-zero 
magnetic moment. 

\section{Discussion and Conclusions}

In this paper we examined the phase effect caused by neutrino magnetic moment in
a core collapse supernova. We assumed that neutrinos are Majorana particles and
have a magnetic moment larger than the Standard Model prediction. The large
magnetic moment shifts the energy eigenstates away from the flavor eigenstates
at the center of the supernova so that each neutrino is emitted as a
superposition of energy eigenstates. We showed that the relative phases
developed by these energy eigenstates create a phase effect (i) if there is a
single partially adiabatic SFP resonance, or (ii) if there are overlapping SFP and MSW
resonances. We argued that this distinguishes SFP phase effect from the ordinary
phase effect because the latter is manifested only when there are two partially
adiabatic MSW resonances. 

We approached the problem both analytically and numerically. Our analytical
approach is based on the assumption of complete decoupling between SFP and MSW
resonances and is applicable only to case (i) above. This approach revealed that
the size of the effect does not simply grow with increasing $\mu B$. Instead, it
depends on an interplay between two factors: The first factor is how far the
initial neutrino is from being a pure energy eigenstate and the second factor is
how far the resonances are from being completely adiabatic or completely
nonadiabatic. Increasing $\mu B$ initially enhances the SFP phase effect due to
the first factor, but increasing $\mu B$ further suppresses it by making SFP
resonance adiabatic. 

Our numerical studies revealed that this simple analytical picture can break
down at late post-bounce times because SFP and MSW resonances start to overlap.
SFP resonance is not universal in the sense that its width in density scale
depends on its physical location. At late post-bounce times both resonances
physically move inward but the SFP resonance also becomes wider in density
scale. Eventually the two resonances start to overlap and the analytical
formulation based on the assumption of decoupled resonances collapses. We showed
that when the overlap happens, SFP phase effect is present even if SFP and MSW
resonances appear to be completely adiabatic from a naive application of the
Landau-Zener approach.  Figs. \ref{random initial for 5e16} and \ref{random
initial for 1e15} demonstrate this effect at late post-bounce times. This is
another aspect that distinguishes SFP phase effect from the ordinary phase
effect which involves universal MSW resonances. 

To discuss and illustrate these details, we first used simple exponential fits
for supernova density profiles at different times. These profiles included  one
SFP and one MSW resonances. For a realistic density distribution some neutrinos
go through more resonances (see Fig.  \ref{fig:baryonProfileShock}), but these
additional resonances can only increase the SFP phase effect.  This is because
once the neutrino is produced, its energy eigenstate components accumulate
relative phases by evolving independently.  If the additional resonances are
partially adiabatic, then they further mix the energy eigenstates and increase the
uncertainties.  If they are not, then the uncertainties remain the same. Also,
the realistic density profile pushes the resonances to the outer regions where
the magnetic field is weaker which causes the SFP phase effect to appear
for larger neutrino magnetic moments.

Our results for a realistic density distribution and a propagating shock wave
demonstrate that those neutrinos with the same energy and the same initial state
arrive at the detector with randomized final states due to the SFP phase effect.
We argued that observing many neutrinos in an energy bin would average out and
effectively remove this randomization, but observing  only a few neutrinos would
not. As a result, the detector may respond to a somewhat irregular or randomized
neutrino signal. We carried out a sample analysis for the capture reactions in
Ar nuclei in DUNE. Our analysis showed that, before the finite energy resolution
of the detector is taken into account, event rate fluctuations due to the SFP
phase effect appear earlier than those due to the ordinary phase effect. Event
rate fluctuations also appear only for those neutrinos with energy larger than
about $25$ MeV, despite the presence of a sizable SFP phase effect below this
energy. When the finite energy resolution of the detector is taken into account,
however, all fluctuations are erased in the energy spectrum of the observed
charged leptons. 

There are some aspects of the problem which are left unexplored. For example,
the electron fraction is taken to be constant at $Y_e=0.45$ throughout the
paper. In the limit $Y_e\to 0.5$ the SFP resonance moves significantly near the
neutrinosphere and this can potentially change our conclusions regarding the
detector response. The fluctuations are erased by the detector resolution
because the relative phase acquired by the energy eigenstates from production to
SFP resonance point is very large and therefore change rapidly with neutrino
energy (see Eq. (\ref{phi_B})). If SFP resonance comes closer to the
neutrinosphere, it is possible that the relative phase becomes small and vary
less with energy to create an observable effect.  We believe that a systematic
exploration of the dependence of SFP phase effect on the amount of overlap
between SFP and MSW resonances is also important for the same reason.  Another
dimension that we did not explore is the effect of neutrino-neutrino
interactions near the proto-neutron star. Neutrino-neutrino interaction
potential creates additional resonances near the neutrinosphere which give rise
to spectral splits \cite{Raffelt:2007cb,Raffelt:2007xt}. If these resonances are
partially adiabatic \cite{Ekinci:2021miy}, then they might have an effect on the
energy dependence of the SFP phase as well. We leave these aspects of the
problem to future publications.

\section*{Acknowledgements}

\noindent T. B. acknowledges the 2214A travel fellowship from the Scientific and
Technological Research Council of Turkey (T{\"{U}}B{\.{I}}TAK) and thanks the
GSI Helmholtz Centre for Heavy Ion Research for their hospitality, where part of
this work was carried out.  Y.P. thanks to the organizors of the workshop on
\emph{Collective Neutrino Oscillations: From Quantum Information Science to
Heavy Element Synthesis} at the Mainz Institute for Theoretical Physics,
Johannes Gutenberg University, where she found an opportunity to discuss this
work with many colleagues.  Numerical calculations reported in this paper were
partially performed at T{\"{U}}B{\.{I}}TAK ULAKB{\.{I}}M High Performance and
Grid Computing Center (TRUBA resources).  This work was supported in part by a
grant from Mimar Sinan Fine Arts University under project number 2018/48. 
\appendix

\bibliographystyle{elsarticle-num} 
\bibliography{TB_nuMag}

\end{document}